\documentclass[%
 reprint,
 amsmath,amssymb,
 aps,
]{revtex4-1}

\usepackage{graphicx}
\usepackage{dcolumn}
\usepackage{bm}
\usepackage[dvipsnames]{xcolor}
\usepackage{multirow}
\usepackage{xr}
\usepackage{hyperref}
\hypersetup{colorlinks=true,linkcolor=black,filecolor=black,urlcolor=black,citecolor=black}
\usepackage{amsmath}
\usepackage{caption}

\newcommand{\be}{\begin{equation}}
\newcommand{\ee}{\end{equation}}
\newcommand{\bs}{\begin{equation} \begin{split}}
\newcommand{\es}{\end{split} \end{equation}}

\begin{document}
\preprint{APS/123-QED}

\title{Responses to transient perturbation can distinguish intrinsic from latent criticality in spiking neural populations}

\author{Jacob T. Crosser$^{1}$}
\author{Braden A. W. Brinkman$^{2,1}$}%
 \email{Corresponding author: braden.brinkman@stonybrook.edu}
\affiliation{$^1$Department of Applied Mathematics and Statistics, Stony Brook University, Stony Brook, NY, 11794, USA}
\affiliation{$^2$Department of Neurobiology and Behavior, Stony Brook University, Stony Brook, NY, 11794, USA}

\date{\today}

\begin{abstract}
The critical brain hypothesis posits that neural circuitry operates near criticality to reap the computational benefits of accessing a wide range of timescales. 
The theory of critical phenomena generally predicts heavy-tailed (power-law) correlations in space and time near criticality, but it has been argued that in the brain such correlations could be inherited from ``latent variables,'' such as external sensory signals that are not directly observed when recording from neural circuitry. 
Distinguishing whether heavy-tailed correlations in neural activity are intrinsically generated within a neural circuit or are driven by unobserved latent variables is crucial for properly interpreting circuit functions.
We argue that measuring neural responses to sudden perturbative inputs, rather than correlations in ongoing activity, can disambiguate these cases.
We demonstrate this approach in a model of stochastic spiking neuron populations receiving external latent input that can be tuned to a critical state.
We propose a scaling theory for the covariance and response functions of the spiking network, which we validate with simulations.
We end by discussing how our approach might generalize to models of neural populations with more realistic biophysical details. 
\end{abstract}

\maketitle


\begin{figure*}[t]
\includegraphics[width=0.875\linewidth]{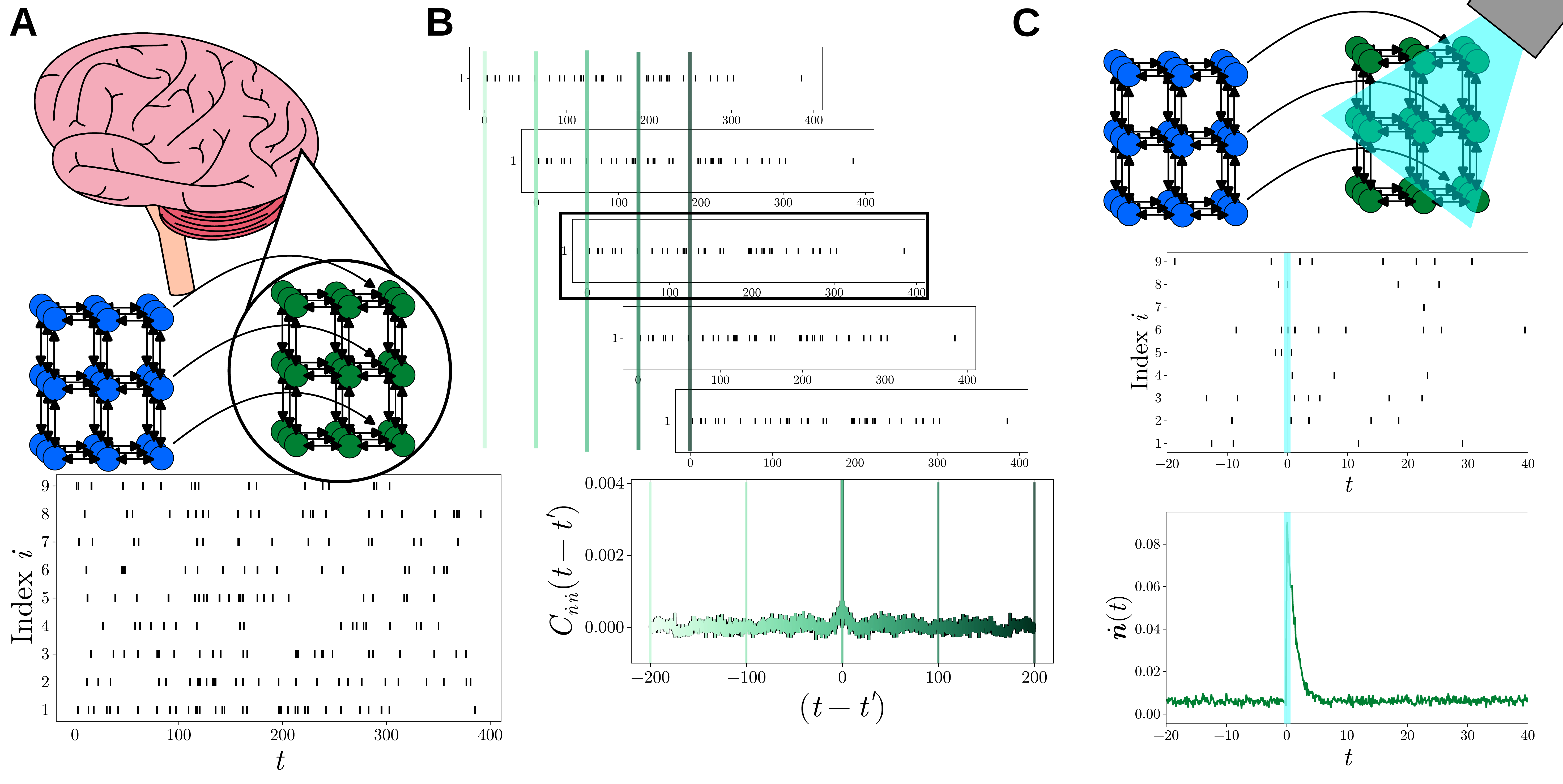}
\caption{\label{fig:Schematic} \textbf{Spiking network model} (A) A network of external inputs (blue) provides feedforward projections to the cortical network (green), which in turn generate discrete spikes (bottom). Inputs are correlated in space and time, representing spatiotemporal correlations in sensory input. The cortical neurons are recurrently connected, leading to spatiotemporal correlations due to both intrinsic circuit connections but also the correlations of the input. (B) The autocovariance of the spike train is given by the expectation of fluctuations around the mean at times $t,\,t'$: $C_{\dot{n}\dot{n}}(t) :=\mathbb{E}[\dot{n}(t+t_0)-\bar{n})(\dot{n}(t_0)-\bar{n})]$. The autocovariance $C_{\dot{n}\dot{n}}$ measures how similar a signal in time (dashed box) is to itself after after shifting the signal by $t$. The superimposed vertical lines show the shifting of the spike trains and the corresponding part of the covariance being calculated. (C) The response function $R_{\dot{n}V}(t)$ of the network is measured by perturbing the steady-state membrane potential by $\Delta V=5.0$ at a time $t_0$ (vertical cyan line) and measuring  the average spike rate at a time $t+t_0$. When the network is an a steady state the covariances and a perturbation from the steady state are functions only of the time difference $t$.}
\end{figure*}

The critical brain hypothesis asserts that networks of neurons in the brain operate near a phase transitions between network states, where the different phases are characterized by the collective activity statistics in the system. 
Near a phase transition, fluctuations span a wide range of spatial and temporal scales  \cite{domb_critical_1985, goldenfeld_lectures_2018}.
It has been argued that this would benefit neural computation by enabling fast  collective responses to stimuli and robust propagation of information through the network \cite{cocchi_criticality_2017,beggs_addressing_2022}.

Experimental evidence of criticality in neural systems is, however, controversial.
Multiple studies have observed putative signatures of criticality in the brain including:
long temporal correlations in $\alpha$-band activity in human EEG and MEG data \cite{linkenkaer-hansen_long-range_2001};
divergence in the variance of statistical surprise in retinal recordings from rats \cite{mora_dynamical_2015};
information-theoretic and thermodynamic signatures of criticality in retinal activity \cite{kastner_critical_2015,tkacik_thermodynamics_2015};
heavy-tailed distributions of the size and duration of neural avalanches \cite{beggs_neuronal_2003,friedman_universal_2012};
power-law distribution of frequencies in human cortical ECG signals \cite{miller_power-law_2009};
and broad dispersion of correlations in macaque motor cortex \cite{dahmen_second_2019}.
However, several modeling studies have argued that these signatures of critical may be artifacts caused by ``latent variables''---unobserved common input to neurons.
Because long correlation lengths or times are characteristic of criticality, strongly correlated common input to weakly coupled or even independent neurons could display heavy-tailed signatures of criticality
\cite{schwab_zipfs_2014,touboul_power-law_2017}.
This confound is especially relevant for sensory cortices: the power-spectra of natural images \cite{ruderman_statistics_1994,ruderman_origins_1997},  fluctuations in loudness and pitch of speech and music \cite{voss_1fnoise_1975}, and the quiet-time distribution of urban soundscapes \cite{de_sousa_scale-free_2019} all exhibit heavy-tailed decays with spatial or temporal frequencies.

While simple models of latent variable input have been used to cast doubt on whether neural circuits themselves are critical, few studies have investigated whether it is possible to disambiguate signatures of criticality that arise due to latent input from those generated intrinsic synaptic coupling between neurons.
Ref.~\cite{ngampruetikorn_extrinsic_2025} investigates the possibility in a generic Ising-like model, finding that mutual information---a difficult quantity to measure in practice---can distinguish these two sources of criticality. 
In this work we propose a complementary scenario for spiking neurons: intrinsic and latent sources of criticality can be distinguished by investigating the transient response of a circuit to large perturbations that excite or suppress the network as a whole, and observing the relaxation to baseline. 
If the network is intrinsically close to criticality, the decay to baseline will be heavy-tailed, independent of any common input to the network.

\textit{Model.--} We demonstrate our claim using a stochastic model of neuron spiking---a nonlinear Hawkes process \cite{truccolo_point_2016,ocker_linking_2017,brinkman_phase_2025,crosser_applications_2024}---that has previously been shown to exhibit criticality \cite{brinkman_phase_2025}. In this work we drive this network strongly correlated external latent inputs.
The dynamics of the membrane potential $V_i(t)$ and spike train $\dot{n}_i(t)$ of the $i$th neuron at time $t$ obeys
\begin{align}
    \tau\frac{d V_i(t)}{dt} = &-V_i(t)+\mathcal{E}\nonumber\\
    & ~~~~+J\sum_{j=1}^N\left(w_{ij}-\delta_{ij}\right)\dot{n}_{j}(t) +x_i(t),\label{eqn:HawkesMembr}\\
\dot{n}_i(t)dt &\sim {\rm Poiss}[\phi(V_i(t))dt],\label{eqn:HawkesSpk}
\end{align}
where $\tau$ is the membrane time constant, $\mathcal E$ is the equilibrium potential of the membrane potentials in the absence of input, $N$ is the number of neurons, $w_{ij}$ is the adjacency matrix, which encodes the connectivity between neurons, $J$ is the magnitude of each synaptic connection, and $x_i(t)$ is the latent input to the neuron, which has dynamics defined in Eq.~\ref{eqn:phi4}.
Neurons fire spikes stochastically at an expected rate $\phi(V_i(t))$ that is conditional on the current membrane potential.
As we are modeling \emph{in vivo} circuitry within the brain, we choose the nonlinear transfer function to be a logistic sigmoid function, $\phi(V) = \left(1 +\exp(-V) \right)^{-1}$, following \cite{brinkman_phase_2025}. 
When a neuron fires a spike, its membrane potential is reset by a fixed amount $-J$, rather than to a fixed value.
In the absence of the external input $x_i(t)$, this spiking network model belongs to the Ising model universality class \cite{brinkman_phase_2025}.

The latent input mimics external signals coming from the sensory periphery, possibly after filtering by upstream circuitry.
We aim to study generic impacts of external input on the spiking network's statistical dynamics, so we model it by a nonlinear stochastic process that can be tuned to a critical point with heavy-tailed spatiotemporal statistics.
We assume $N$ effective latent processes $x_i(t)$, one for each neuron in the network, which obey the dynamics
\begin{equation}
    \tau \frac{dx_i(t)}{dt}=-gx_i(t)^3+\sum_{j=1}^N\left( {w}_{ij}-r\delta_{ij}\right)x_j(t) + \eta_i(t),
    \label{eqn:phi4}
\end{equation}
where $x_i(t)$ is the activity of latent unit $i$ at time $t$, $r$ is the strength of the self-modulation of the unit's own activity, $g$ is the strength of the nonlinear relaxation, $w_{ij}$ encodes the connections between units, and $\eta_i(t)$ is a Gaussian white noise process with mean $\langle \eta_i(t) \rangle = 0$ and covariance $\langle \eta_i(t) \eta_j(t') \rangle = 2\sigma^2\delta_{ij} \delta(t-t')$, where $\sqrt{2\sigma^2}$ is the strength of the noise.
Eq.~\ref{eqn:phi4} corresponds to the so-called ``Model A'' on a lattice, a non-equilibrium extension of the $\phi^4$ model that is also in the Ising model universality class \cite{tauber_field_2007}: 
by tuning the self-coupling $r$, the input network transitions from a state of mean-zero activity, $\langle x_i(t) \rangle = 0$, to a state with $\langle x_i(t) \rangle \neq 0$.

For simplicity, we choose the organization of the input network, $\mathbf{w}$, to match the arrangement of neurons in the spiking network, such that each input unit provides direct input to one neuron in the spiking network. 
We model $\mathbf{w}$ as the adjacency matrix for a hypercubic lattice with periodic boundary conditions, such that $w_{ij} = 1$ if neurons $i$ and $j$ are nearest neighbors and $0$ otherwise.
In $d$ dimensions each neuron has $2d$ neighbors, not including the external input connection.
We focus on a three-dimensional lattice so that we can exploit known properties of the Ising model universality class to facilitate analyses of the combined network at criticality.
However, because real neural circuits are not arranged in neat crystalline lattices, we will focus on detecting signatures of criticality in the temporal activity of the network.
We adapt the scaling theory of Ref.~\cite{brinkman_phase_2025} to model the nonlinear response  of the network to a wide-field excitation or suppression of the network activity, and contrast this with a scaling theory we develop for the spike-spike covariances.

\label{sec:scaling}
\textit{Scaling theory--}
We expect the tails of the decay of the population response to obey the same scaling form for the relaxation of the population mean studied in \cite{brinkman_phase_2025},
\begin{align}
    R_\pm(t)-\nu_c &\sim t^{-\frac{d-2+\eta_\ast}{2 z_\ast}} F_\pm(t/\xi_{\rm spk}),
    \label{eqn:responsescaling}
\end{align}
where $\nu_c$ is the baseline population-averaged firing rate of the network at the critical point, $F_\pm$ is a scaling function of the ratio of the time $t$ to a coherence time $\xi_{\rm spk}$, $d$ is the dimension of the lattice, $\eta_\ast$ is the ``anomalous'' critical exponent and $z_\ast$ is the ``dynamic'' critical exponent. 
The subscript $+$ indicates the response to an excitation of the network and $-$ indices the response to a suppression of the network.
In practice, accurately estimating $\nu_c$ may be difficult, so we consider the difference in responses, $R_{\dot{n}V}(t) \equiv R_+(t) - R_-(t)$, which eliminates $\nu_c$ and obeys a similar scaling form, $R_{\dot{n}V}(t) \sim t^{-\frac{d-2+\eta_\ast}{ 2 z_\ast}} \bar{F}(t/\xi_{\rm spk})$.
The coherence time $\xi_{\rm spk}$ scales with the distance to the critical point. 
Following \cite{brinkman_phase_2025}, we focus on the case in which the network is tuned to its critical synaptic strength $J_c$ and we tune the input $\mathcal E$. 
In this case, $\xi_{\rm spk} \sim (\mathcal E - \mathcal E_c)^{-\frac{2z_\ast}{d+2-\eta_\ast}}$.

Because the response of the network to this perturbation is causally independent of the latent input activity, the response function itself does not depend on the state of the latent input process.
In contrast, the spike-spike covariances do depend on the latent input.

\begin{figure*}
\includegraphics[width=\linewidth]{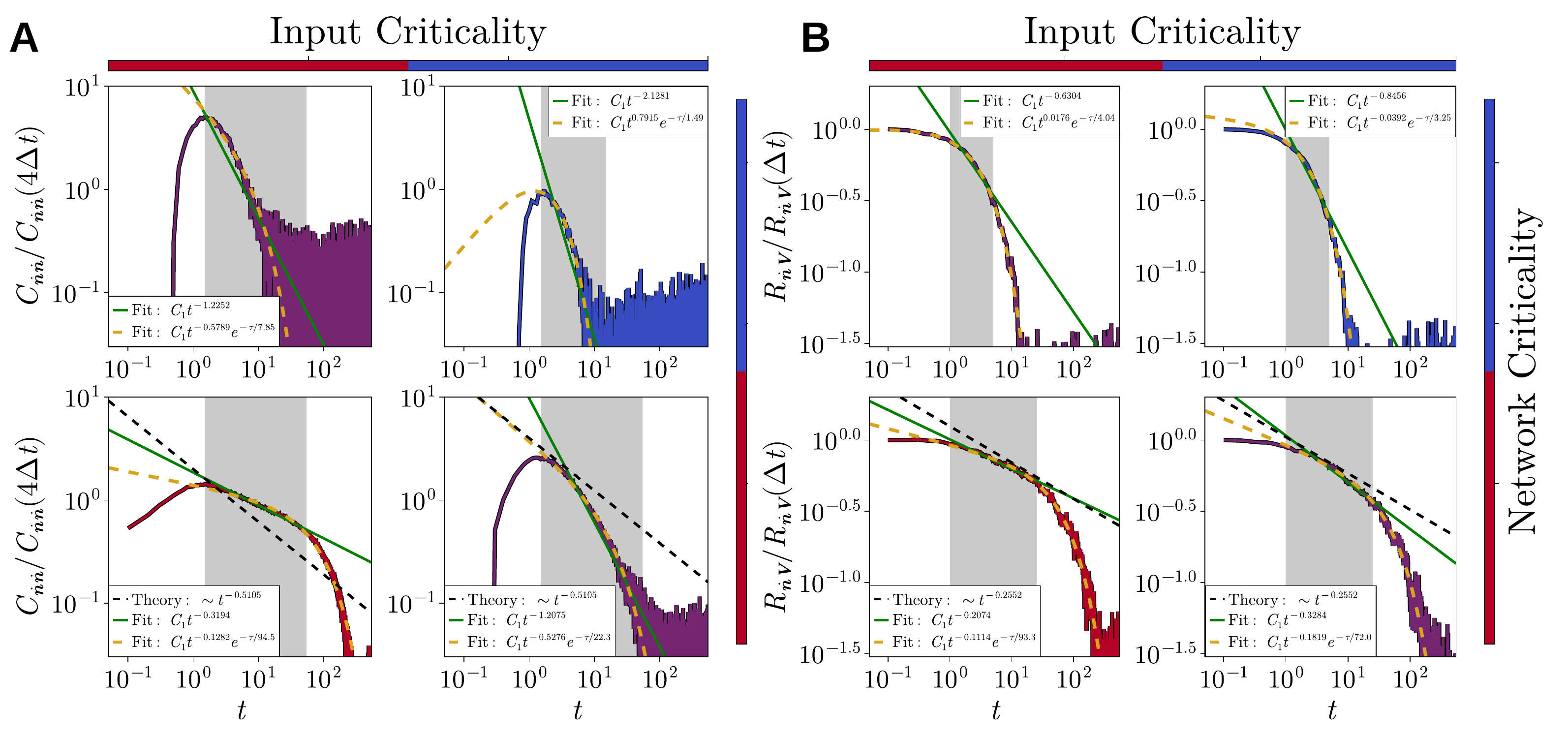}
\caption{\label{fig:SpkGrid} \textbf{Spike covariance and response functions in a 25$^3$ lattice}
Each color bar denotes whether the given subsystem is close to criticality (red) or far from criticality (blue).
(A) Spike-spike covariance functions corresponding to the combinations of the input and spiking network being near or far from their respective critical points.
(B) The differential response of the average spike rate in response to a perturbation of size $\Delta V = \pm5$ to the membrane potential of each neuron.
Dark gray boxes show the region over which the power law (green) was estimated; fits of the power-law exponential (gold) start from the left edge of the gray box and continue until the end of the trial.
Spiking network: far-from-critical (blue) with $\xi_\text{spk}= 0.5$, near-critical (red) with $J_c - J=2\times10^{-4}$.
Input network: far-from-critical (blue) with $r - r_c =1.0$, near-critical (red) with $r-r_c=10^{-3}$.}
\end{figure*}

Accurately estimating the scaling form of the population-averaged autocovariance is more challenging.
In the absence of input, we expect $C_{\dot{n}\dot{n}}(t) \sim t^{-(d-2+\eta)/z} G(t/\xi_{\rm spk})$, which follows from the general spatiotemporal scaling form for non-equilibrium critical phenomena \cite{tauber_field_2007}.
In $d = 3$ we expect $\eta_\ast \approx 0.036$ and $z_\ast \approx 2.02$, which are close to the mean-field values $\eta_\ast = 0$ and $z_\ast = 2$. 
We therefore estimate the scaling of the covariance using a mean-field approximation, which yields
\begin{align}
C_{\dot{n}\dot{n}}(t) &\sim t^{-d/2+1} G(t/\xi_{\rm spk}) \nonumber\\
& ~~~~~ + \sigma^2 t^{-d/2+1}{\rm min}(\xi_{\rm spk},\xi_{\rm lat})^2\mathcal G(t/\xi_{\rm spk}),
\label{eqn:Cnn-full_MFScaling}
\end{align}
where we introduce the coherence time of the latent input, $\xi_{\rm lat}^{-1} = (r_c - r)/\tau$.
The second term in this expression arises from the latent input process.
The limiting behavior of the scaling function $\mathcal G$ depends on the size of the ratio $\xi_{\rm spk}/\xi_{\rm lat}$, not denoted explicitly. 
If $\xi_{\rm spk} \ll \xi_{\rm lat}$, then $\mathcal G$ is constant to leading order, while it exhibits exponential decay if $\xi_{\rm spk} \gtrsim \xi_{\rm lat}$. 
As a result, both the intrinsic spike-spike covariance and the covariance induced by the latent population are predicted to have similar temporal dynamics and scaling forms---at least when one of the spiking network or latent input process are away from their critical point.
This mean-field prediction breaks down if both the network and input are critical, and therefore may not even make good qualitative predictions in the doubly-critical regime.
This is in contrast to our simulations, which appear to give well-behaved results (Fig.~\ref{fig:SpkGrid}A, bottom left).

To verify the qualitative predictions of our scaling theory, we turn to simulations.

\textit{Simulations and scaling collapses--}
We simulate the full set of coupled stochastic differential equations on a $25^3$ lattice with a timestep of $\Delta t=0.1$~ms. 
We first allow the joint system to relax to its steady-state.
To estimate the autocovariance of the network we allow the dynamics to evolve for an additional $15$~s of simulation time.
To estimate response functions, we draw initial conditions from a $750$~ms after the system equilibrates, boosting or suppressing the membrane potential of the neurons by $\Delta V = \pm 5$ and allowing the network to relax back to steady-state for $15$~s. 
We average the recorded autocovariance or responses across a population of 25 spatially adjacent neurons and across 10 separate trials for covariance estimates and $400$ trials for response estimates.
Subsampling $25$ out of the $25^3$ neurons mimics the experimental limitation of not being about to record from every neuron in the network, and allows us to limit the number of pairwise covariances we must calculate, which is computationally expensive to do for the entire network.

We compare the covariance and responses in Fig.~\ref{fig:SpkGrid} for four cases: sub- or near-critical latent input and sub- or near-critical neural population.
When at least one of the networks is critical, we expect the covariances to exhibit power-law decay, while we only expect power laws in the response functions if the neural population is critical.
Our simulations confirm our expectations, though we find that these power laws are significantly cut off by finite size effects that distort the apparent exponent of the intermediate heavy-tailed regime.
However, accounting for an exponential cutoff yields estimates consistent with the Ising model universality class.

\begin{figure*}[t]
\includegraphics[width=\linewidth]{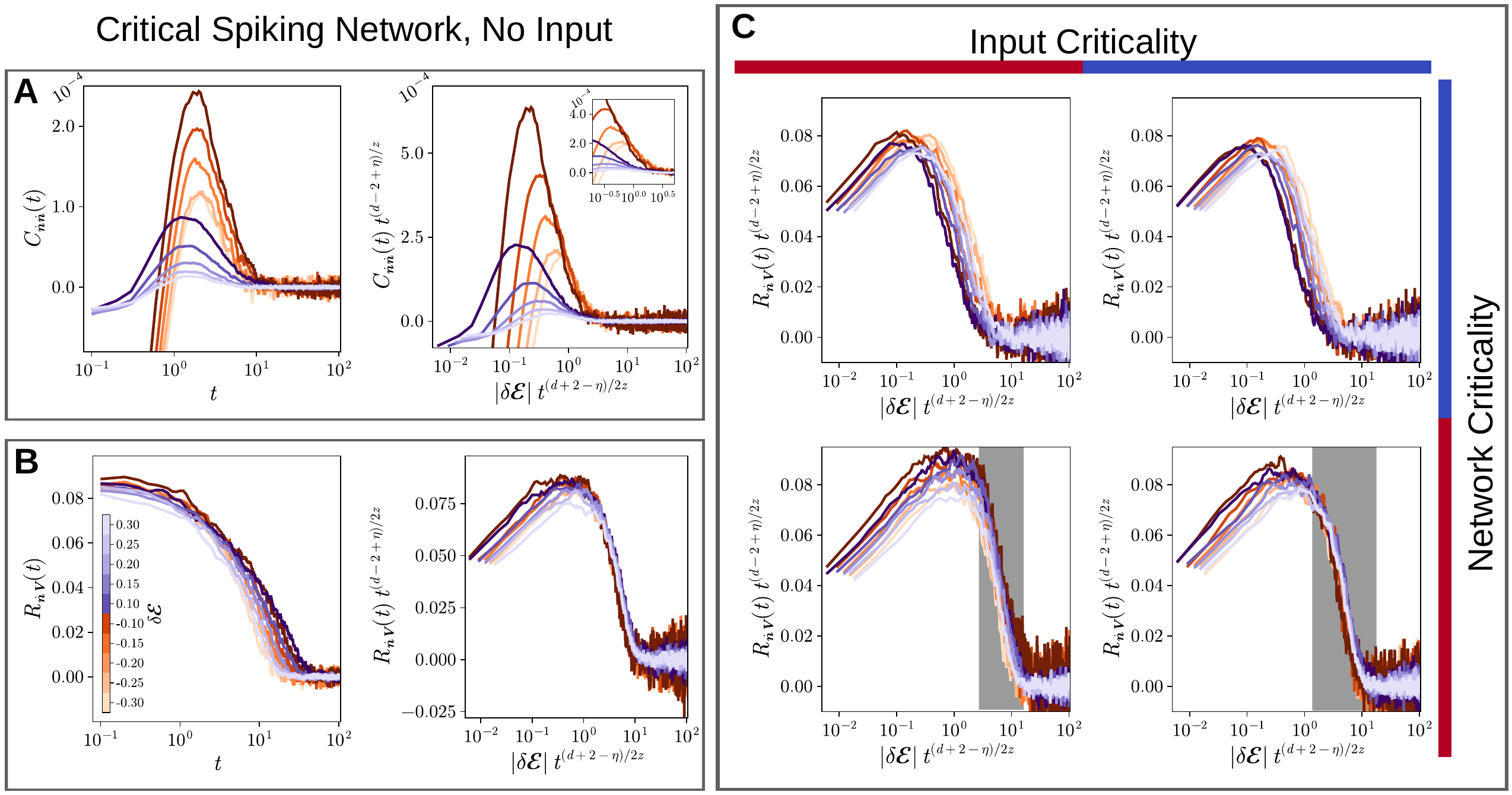}
\caption{\label{fig:SpkCollapse} \textbf{Shape collapse of the covariance and response functions in a 25$^3$ lattice} (A) Collapse of the spike-spike autocovariance in a network at criticality and without any input. (B) Shape collapse of the spike response functions in the same conditions as in (A). (C) Collapse of the spike response functions to a $\Delta V=5.0$ perturbation from the conditions in Fig.~\ref{fig:SpkGrid}.}
\end{figure*}

A more robust way to estimate critical exponents is to perform a data collapse using the predicted scaling forms (\ref{eqn:responsescaling}) and (\ref{eqn:Cnn-full_MFScaling}).
As we will see, collapses of our simulated data appear to be less sensitive to finite-size effects.
Fig.~\ref{fig:SpkCollapse}A shows the collapse of the covariance function for a network at its critical point and without any input ($\sigma = 0$), while Fig.~\ref{fig:SpkCollapse}B depicts the collapse of the spike response function under the same conditions.
In the presence of latent input, the quality of the shape collapse is strongly dependent on the distance of the network from its critical point, and more weakly dependent on the state of the latent input. Fig.~\ref{fig:SpkCollapse}C shows four such attempted collapses when the latent and spiking networks are close or far from their respective critical points by tuning $r$ and $J$.
The collapses are then performed over networks tuned away from the critical value $\mathcal E_c(J_c)$.
The collapse is successful when the network is close to $J_c$ (bottom row), and less successful when it is not (top row), to be expected because the network is far from the critical point in both $J$ and $\mathcal E$. 
Although the state of the latent input network does not modify the scaling form of the nonlinear response function, it does impact the quality of the collapse because the input fluctuations are stronger when the latent network is near its critical point.
More trials would be required to further average down the variability caused by the latent input.

\textit{Discussion--}
Numerous studies have proposed that signatures of criticality in neural systems may be artifacts arising from independent or weakly connected neurons being driven by common latent (unobserved) variables \cite{schwab_zipfs_2014,touboul_power-law_2017,morrell_latent_2021,morrell_neural_2024}.
This, in our view, is not a useful way to think about criticality in neural systems.
The Ising model---\emph{the} prototypical example of criticality---can be written as a collection of independent spins coupled to a set of unobserved latent variables.
When those latent variables are averaged out the well-known form of the Ising model is recovered, as pointed out in Ref.~\cite{meshulam_statistical_2025}.
This said, it is still a valid and important question to ask whether signatures of criticality are locally generated or inherited from upstream unobserved sources.
The answer to this question informs where we should focus experimental resources if we seek to manipulate criticality in neural systems.

Distinguishing the effects of recurrent synaptic connections from common input has been a long-standing challenge in neuroscience \cite{ngampruetikorn_extrinsic_2025,kulkarni_common-input_2007,farina_common_2015,rodriguez-falces_correlation_2017,castronovo_decrease_2018}.
Most statistical analyses of neural data focus on covariances in neural activity, which cannot generally separate the influences of the input from the internally generated dynamics (Fig.~\ref{fig:SpkGrid}A).
However, if one instead measures the trial-averaged responses of the network to repeated perturbations, the influence of the inputs is averaged out (Fig.~\ref{fig:SpkGrid}B), and the response functions of networks close to a critical point can be collapsed onto universal curves, as demonstrated in Fig.~\ref{fig:SpkCollapse}.

For the sake of analytic tractability, we focused on a simple model in which the the connectivity of both the input network and the spiking network are exactly the same, with a $1$-to-$1$ mapping from input units to neurons.
In reality these networks would be quite different, and multiple inputs could be delivered to single neurons.
Moreover, the synaptic connections would not be organized on a crystalline-like lattice, nor would the strength of the connections be homogeneous, as in the network studied here.
While Ref.~\cite{brinkman_phase_2025} established criticality holds in homogeneous networks without spatial organization, the role of heterogeneity in synaptic connections and its effect on criticality in spiking networks remains an open question.
Such ``disorder'' can have a variety of effects on the critical properties of homogeneous systems, ranging from changing the values of the critical exponents to smearing out the phase transition, effectively destroying it.
However, in systems in which criticality is preserved we expect our main result to qualitatively hold: response functions will only display heavy-tailed decays in populations that are intrinsically critical, and response functions are therefore more robust signatures of criticality than correlational data.

\begin{figure*}[t]
\includegraphics[width=0.9\linewidth]{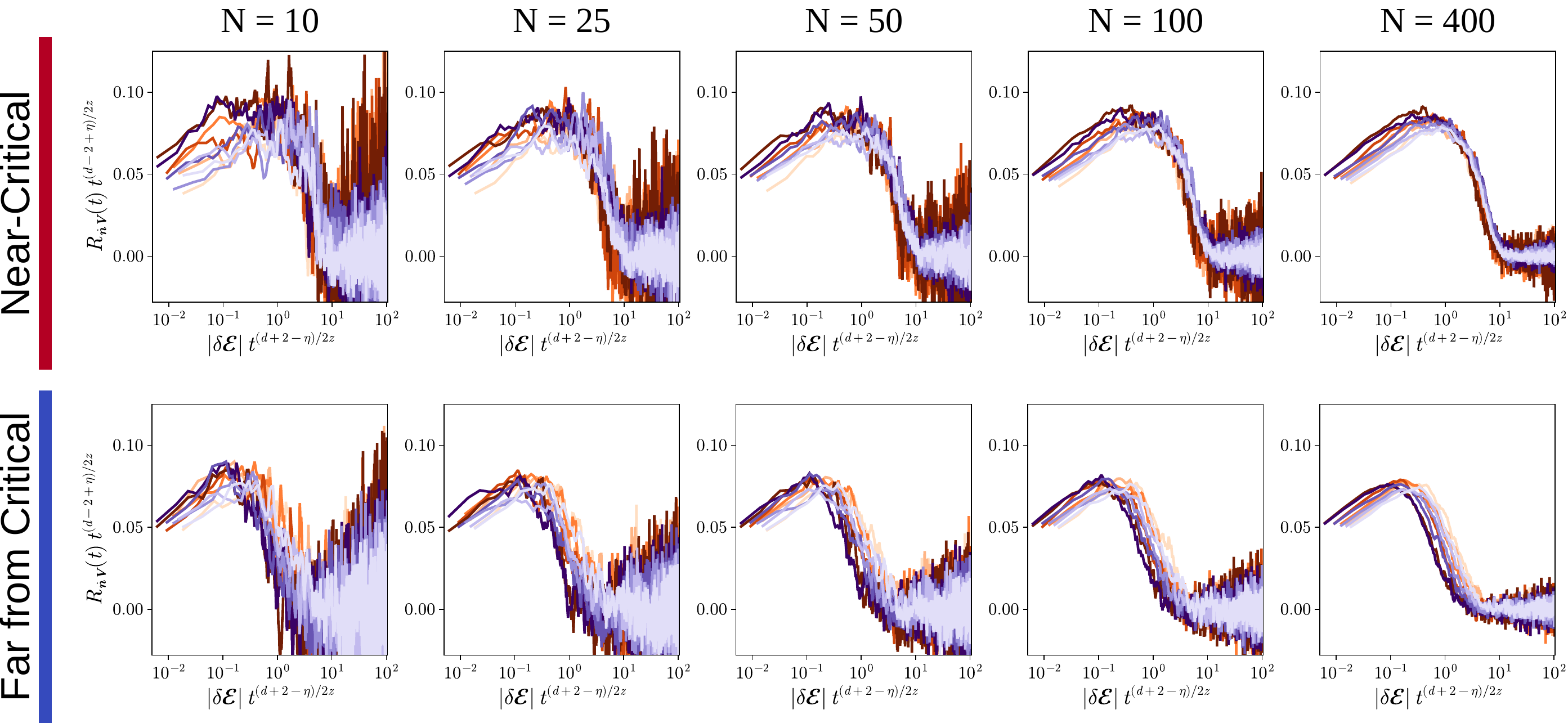}
\caption{\label{fig:SpkCollapseTitr} \textbf{Quality of response collapse with number of trials}
The attempted collapse of response functions averaged over an increasing number of trials for the cases of:
(Top) a near-critical spiking network ($J_c-J =0.0002$);
(Bottom) a far-from-critical spiking ($J_c-J=0.5$).}
\end{figure*}

This said, there are practical challenges to using response functions over correlations.
For one, covariances can be estimated from long recordings of steady-state activity, whereas response functions require perturbations to drive the network out of its steady state and allow it to relax back.
However, if a network is close to a critical point, this relaxation can be very slow, making it difficult to obtain sufficiently many trials to average over.
The quality of the collapse will be reduced with insufficiently many trials; see Fig.~\ref{fig:SpkCollapseTitr} in the End Matter.
It may be more practical to attempt to tune networks away from their critical points so that the relaxation is faster, but still close enough to criticality that scaling collapses can be performed.
Recent experimental work suggests that one means of tuning circuits away from criticality is arousal \cite{fontenele_low-dimensional_2024,xu_sleep_2024}.
New experimental technologies for performing perturbations have also been developed over the past several years, such as optogenetic stimulation,
opening a new frontier of exploring neural response functions and tests of brain criticality.


%


\clearpage
\begin{center}
    \textbf{End matter}
\end{center}
\textit{Phase diagram of the spiking network model}---Phase transitions in the spiking network model have previously been detailed in Ref.~\cite{brinkman_phase_2025}. 
In this section we summarize the phase diagram as obtained by a mean-field approximation and our simulations.
The mean-field approximation amounts to neglecting fluctuations in the spiking activity and the latent input, which amounts to replacing $\dot{n}_j(t)$ with $\phi(V_j(t))$ in Eq.~\ref{eqn:HawkesMembr} and $\eta_i(t)$ with $0$ in Eq.~\ref{eqn:phi4}.

The latent input undergoes a bifurcation as a function of the leak $r$: for $r > r_c$ the mean activity is zero, while for $r < r_c$ the mean input scales as $\langle x \rangle \sim \pm \sqrt{r_c-r}$.
In a $d=3$ toroidal lattice, $r_c = 2d = 6$.

\begin{figure*}[!th]
\includegraphics[width=0.8\linewidth]{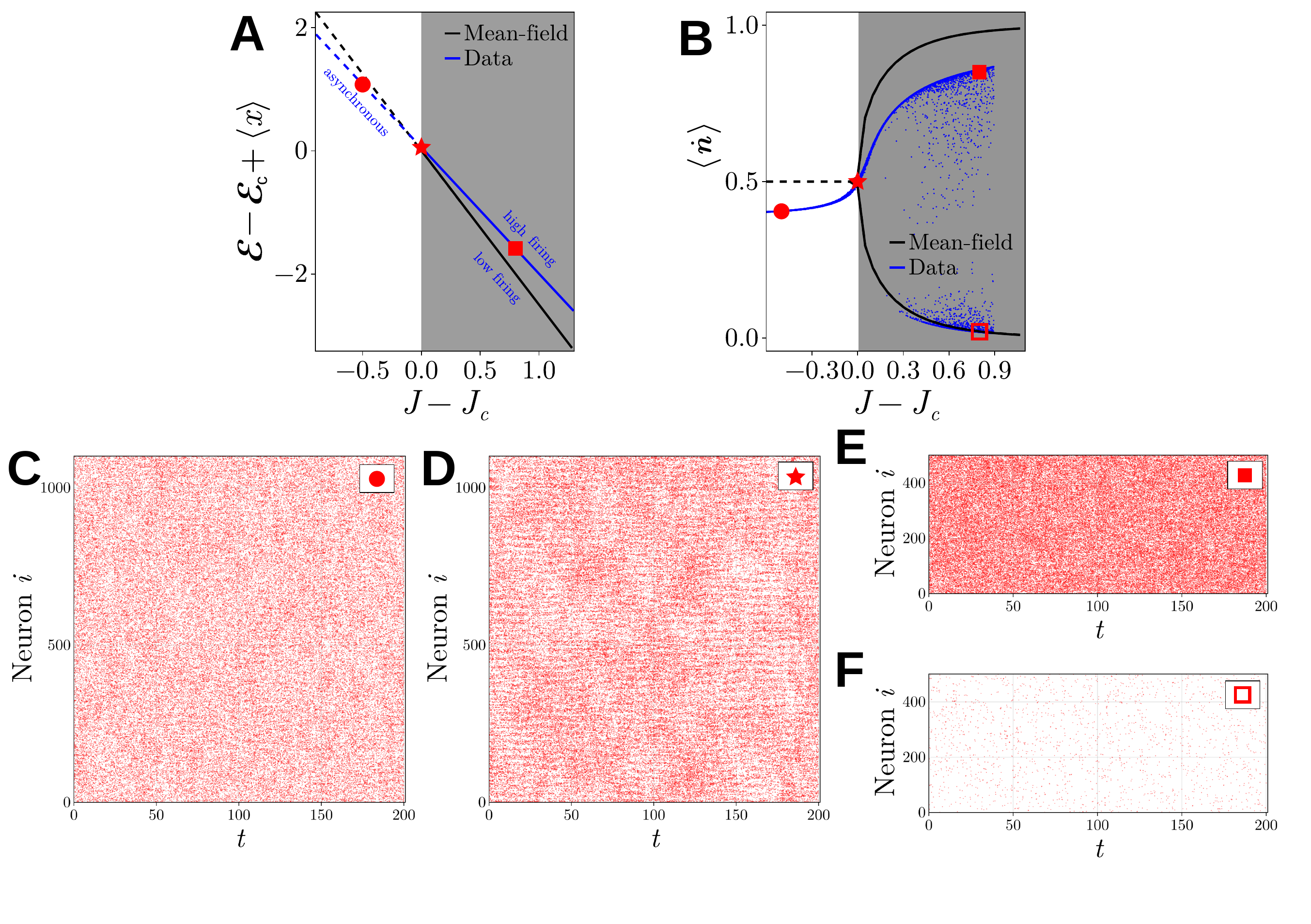}
\caption{\label{fig:PhaseDiagram} \textbf{Phases of neural activity in the spiking network model} (A) Phase diagram of the spiking model outlined in Eqs.~(\ref{eqn:HawkesMembr}-\ref{eqn:HawkesSpk}) on a 25$^3$ lattice. (B) Average firing rate of the spiking model along the bifurcation ridge in (A). (C)-(F) Example spike event rasters from various points along the bifurcation ridge.}
\end{figure*}

The spiking network with sigmoidal nonlinearity undergoes a similar bifurcation at a critical point $(\mathcal E_c,J_c)$, at which the net tonic (time-independent) input to each neuron is canceled out, and the passive leak of current out of the cell is also balanced by the synaptic input \cite{brinkman_phase_2025}.
This yields the criticality conditions
\begin{align}
\mathcal E_c(J) - \theta + J_c(2d-1)\phi(\theta) + \langle x \rangle &= 0,\nonumber\\
1 - J_c(2d-1)\phi'(\theta) &= 0,\label{eqn:CriticalityConditions}
\end{align}
where $\theta = {\rm argmax} ~\phi'(V)$, which is $0$ for the sigmoidal nonlinearity we consider in this work.
If the latent input network is sub-critical or critical, then $\langle x \rangle = 0$.
We will not focus on the case of supercritical networks in this work, which would induce a non-zero shift of the critical resting potential $\mathcal E_c$.
In $d=3$ and for $\phi(V) = (1+\exp(-V))^{-1}$, the criticality conditions yield the mean-field transition point $(\mathcal E_c, J_c) = (-2,4/5)$ \footnote{Note that we can also observe a second kind of bifurcation in this model that satisfies the criticality conditions for $\theta \neq {\rm argmax}~\phi'(\theta)$ by allowing $J > J_c$ and appropriately tuning $\mathcal E$. This corresponds to a so-called ``spinodal'' transition, in which the network is tuned to the edge of stability of one of the metastable branches of the network's equation of state. However, it has been debated whether such transitions exist in the actual stochastic network. While recent work suggests that hysteresis experiments may provide a means of testing spinodal transitions, this is beyond the scope of this study.}.
When $\mathcal E < \mathcal E_c$ the network activity is primed towards lower firing rates, while $\mathcal E > \mathcal E_c$ induces higher firing rates.
Along the $\mathcal E = \mathcal E_c$ curve we observe a transition similar to the input network: for $J < J_c$ there is a single steady-state with mean firing rate $\langle \dot{n} \rangle = \phi(\theta)$, which splits into low and high firing rate states as $J > J_c$ (Fig.~\ref{fig:PhaseDiagram}A).
The phase diagram in the full stochastic model is qualitatively the same, but the quantitative predictions of the critical line $\mathcal E_c(J)$ and the steady-state firing rates differ due to the nonlinear impact of the stochastic fluctuations \cite{brinkman_phase_2025}.
In Figs.~\ref{fig:SpkGrid} and \ref{fig:SpkCollapse} we vary the ``distance to criticality'' of the spiking network by moving along the estimated critical line $\mathcal E_c(J)$.
If we move too far off the critical line the bias caused by the tonic input $\mathcal E$ will overcome the influence of the latent input, even if the latent input is in its critical state, and would exhibit exponential decay of correlations rather than a heavy-tailed decay.

\textit{Dependence on number of trials}---The collapse results we present in the main section have been averaged over $400$ simulation trials to demonstrate what successful collapses look like under ideal circumstances.
Such a large number of trials may be difficult to achieve in most experiments, although some labs are capable of continuously recording the same neurons in an animal for up to a day \cite{xu_sleep_2024}. 
We therefore show how the number of trials impacts the quality of the response function collapse in Fig.~\ref{fig:SpkCollapseTitr}.
Recall, however, that we limited the number of neurons recorded in our simulated responses to be only $25$ out of $25^3$ neurons; the variability can therefore also be averaged down by averaging over more neurons, rather than more trials.

\clearpage

\newpage


\begin{center}
    \Large\textbf{Supplemental Material}
\end{center}\normalsize
\vspace{-0.5cm}
\appendix 
\section{Basic Methodology}\label{sec:MethodDetails}
\begin{figure*}[t!]
\begin{center}
    \includegraphics[width=0.9\linewidth]{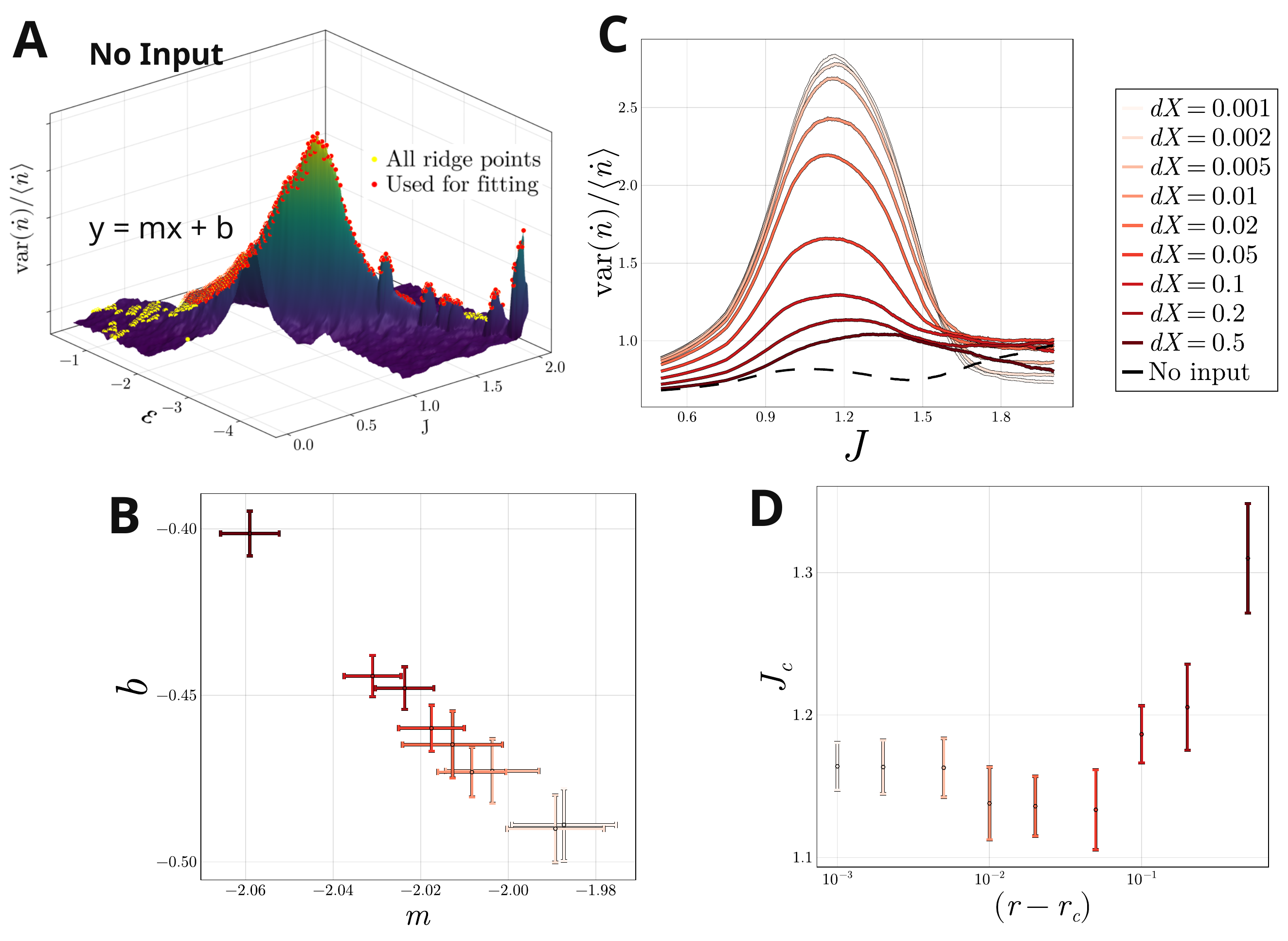}
\end{center}
\caption{\label{fig:HT_RidgeFitting} \textbf{Finding critical points in spiking networks}
A graphical sketch of the procedure used to find estimate the critical ridge $\mathcal{E}_c(J)$ and the critical recurrent weight $J_c$.
(A) The normalized variance of the spike acts as an indicator of criticality.
The values of $\mathcal{E}$ where the normalized network variance were in the 90th percentile and above for a given value of $J$ were identified as possible ridge points (yellow).
Of all possible ridge points, the top 90\% were identified as true ridge points (red) and a line was fit to them.
(B) The slope $m$ and intercept $b$ of the ridge fits plotted parametrically as a function of $(r-r_c)$.
(C) A new set of parameters along these fit lines was run, and the normalized variance was smoothed with either a symmetric boxcar window ($J_c-J = 0.5$).
(D) The critical value $J_c$ is taken to be the value of $J$ at which the variance in (C) first peaks when $J$ increases.}
\end{figure*}

\textit{Simulation conditions}\label{ssec:SimMethods}---Simulations of the spiking network and the $\phi^4$ noise process $\mathbf{x}$ were run using a forward-Euler integration scheme with a time step of $\Delta t=0.1$ in unitless time.
Prior to collecting of simulation data for analysis, the models were initialized with $\mathbf{V}_0= \mathbf{x}_0 = \mathbf{0}$ and allowed to run for $t_{\text{init}}=2000$ units of time to relax to a stationary distribution of activity, noting that the initial values $\mathbf{V}_0$ and $\mathbf{x}_0$ correspond to the mean field stationary point in their respective subcritical regimes.
Simulations used for scanning the $(J,\mathcal{E})$ parameter plane of the spiking networks (Fig.~\ref{fig:HT_RidgeFitting}A) were initiated after another $3000$ units of time (for all $n_\text{trials}=50$) and data was collected for the subsequent $500$ time units.

Simulations used to calculate the population-averaged covariance and response functions underwent a process of multi-point initialization.
After relaxing to a stationary state for $t_{\text{init}}=2000$ units of time, the initializing simulation was allowed run for another $1000$ time units from which $n_\text{trials}$ evenly spread network snapshots were taken. These snapshots were used as initial conditions for subsequent data-collecting trial simulations.
Networks were simulated for $t_{\text{max}}=50,000$ to estimate the autocovariance functions before averaging across $n_\text{trials}=5$ trials.
When calculating the response functions, networks were simulated for $t_\text{max}=1000$ time units and averaged across $n_\text{trials}=400$ trials.
In both the covariance and response function calculations, the activity of 25 neurons---1 central neuron, its nearest neighbors, and next-nearest neighbors---over which the functions are averaged.

\textit{Statistical measures}\label{ssec:StatMethods}---
Here, we will outline the particularities of statistical measures reported in this manuscript.
The mean values for the spike train reported in Fig.~\ref{fig:HT_RidgeFitting} are formally calculated as
\begin{equation}
    |\langle \dot{n}\rangle| := \langle \left|\langle \langle \dot{n} \rangle_\text{pop}\rangle_\text{time}\right|\rangle_\text{trial},\label{eqn:Phi_mean}
\end{equation}
The spike train variance shown in blue in Fig.~\ref{fig:HT_RidgeFitting} is given by 
\begin{equation}
    \text{var}(\langle \dot{n}\rangle):=\langle \text{var}_\text{time}( \langle \dot{n} \rangle_\text{pop})\rangle_\text{trial},
\end{equation}
which is analogous to the trial-averaged susceptibility of the magnetization used to describe Ising models the statistical mechanics literature.
The angle brackets denote an average over one of \{population, time, trials\}, and expectations are taken in order from innermost to outermost.

\section{Path integral formalism and the Gaussian process approximation}\label{app:PathIntGPA}

To model the spiking dynamics of individual neurons, we consider a nonlinear Hawkes process \cite{ocker_linking_2017,brinkman_predicting_2018} introduced in Eqs.~(\ref{eqn:HawkesMembr}-\ref{eqn:HawkesSpk}) together with the latent input process modeled Eq.~(\ref{eqn:phi4}).

As in previous work \cite{chow_path_2015,ocker_linking_2017,crosser_applications_2024,brinkman_phase_2025}, we leverage a path integral formalism to approximately calculate the statistical properties of the network activity. 
The joint distribution for the whole system can be written as 
\begin{align*}
    P\left[\mathbf{V}(t),\mathbf{\dot{n}}(t),\mathbf{x}(t)\right]~&=\int\mathfrak{D}\left[\mathbf{\tilde{V}},\mathbf{\tilde{n}},\mathbf{\tilde{x}}\right]~e^{-S\left[\mathbf{\tilde{V}},\mathbf{V},\mathbf{\tilde{n}},\mathbf{\dot{n}},\mathbf{\tilde{x}},\mathbf{x}\right]},
\end{align*}
where we have introduced the auxiliary response variables $\mathbf{\tilde{V}}(t)$, $\mathbf{\tilde{n}}(t)$, and $\mathbf{\tilde{x}}(t)$. 
The joint action, $S\left[\mathbf{\tilde{V}},\mathbf{V},\mathbf{\tilde{n}},\mathbf{\dot{n}},\mathbf{\tilde{x}},\mathbf{x}\right]$, can be rewritten as 
\begin{equation*}
S\left[\mathbf{\tilde{V}},\mathbf{V},\mathbf{\tilde{n}},\mathbf{\dot{n}},\mathbf{\tilde{x}},\mathbf{x}\right]=S_{\rm net}\left[\mathbf{\tilde{V}},\mathbf{V},\mathbf{\tilde{n}},\mathbf{\dot{n}}|\mathbf{x}\right]+S_{\rm input}\left[\mathbf{\tilde{x}},\mathbf{x}\right].
\end{equation*}
With this, we can understand the total action to be the sum of the network action conditioned on the external noise and the action for the external noise itself. 
The noise-conditioned action for the network is given by
\begin{widetext}
    \begin{align*}
        &S_{\rm net}\left[\mathbf{\tilde{V}},\mathbf{V},\mathbf{\tilde{n}},\mathbf{\dot{n}}|\mathbf{x}\right]\nonumber\\
        &~~=\int dt~\sum_{i=1}^N\Bigg\{\tilde{V}_i(t)\Bigg[\dot{V}_i(t)+V_i(t)-\mathcal{E}_i-x_i(t) -J\sum_{j=1}^N\left(w_{ij}-\delta_{ij}\right)\dot{n}_j(t)\Bigg] +\tilde{n}_i(t)\dot{n}_i(t) -\left(e^{\tilde{n}_i(t)}-1\right)\phi(V_i(t)) \Bigg\}. 
    \end{align*}
\end{widetext}

In contrast to the approach in \cite{crosser_applications_2024},
we retain all variable pairs in $S[\tilde{\mathbf{V}},\mathbf{V},\tilde{\mathbf{n}},\dot{\mathbf{n}},\tilde{\mathbf{x}},\mathbf{x}]$ in the present work to understand the statistics of both the membrane potentials $\mathbf{V}$ and the spiking processes $\dot{\mathbf{n}}$.
Additionally, here we study the spatiotemporal covariance functions $ {C}_{i,j}^{P,Q}(\tau)$ and response functions $ {R}_{i,j}^{P\rightarrow Q}(\tau)$, where $P,Q\in\{\mathbf{V},\dot{\mathbf{n}},\mathbf{x}\}$; this is in contrast to the previous manuscript where we restricted ourselves to the static covariance matrix of the membrane potential $ {C}_{i,j}^{\mathbf{V},\mathbf{V}}:= {C}_{i,j}^{\mathbf{V},\mathbf{V}}(\tau)|_{\tau=0}$. 
The full list of covariance and response functions under this second-order approximation are derived in Appendix \ref{app:PathIntGPA} and given in full in Appendix \ref{app:rcf_GPA}, though we will discuss some of those results in the main body where relevant.

The action governing the dynamics of the latent input $\mathbf{x}(t)$ is 
\begin{widetext}
\begin{align*}
S\left[\mathbf{\tilde{x}},\mathbf{x} \right] &= \int dt~\sum_{i=1}^N\Bigg\{\tilde{x}_i(t)\left[\dot{x}_i+\sum_{j=1}\left( {w}_{ij}^{x}-r\delta_{ij}\right)x_j+gx_i^3\right] + \sigma^2 \tilde{x}_i^2(t) \Bigg\},
\end{align*}
\end{widetext}
which corresponds to a stochastic process 
\begin{equation*}
    \begin{split}
        \dot{x}_i(t) = & -\sum_{j=1}\left( {w}_{ij}^{x}+r\delta_{ij}\right)x_j(t)+gx_i^3(t)+\sqrt{2\sigma^2}\eta_i(t),
    \end{split}  
\end{equation*}
where $\eta_i(t)$ is a zero-mean Gaussian process with covariance $\langle \eta_i(t) \eta_j(t')\rangle = 2\sigma^2 \delta(t-t')$.
In the action $S\left[\mathbf{\tilde{x}},\mathbf{x} \right]$ the noise has been marginalized over to give a functional of $\mathbf{x}$ and $\mathbf{\tilde{x}}$.

An important step in the current analysis of network correlations is to take a second-order (``saddle-point'') approximation of the action $S$ around the mean-field solution to the dynamics.
Because the leading mode of the adjacency matrix $\mathbf{w}$ id homogeneous, the dynamics of the average membrane potential follows 
\begin{equation*}
    \frac{d\langle V\rangle}{dt} = -(\langle V\rangle-\mathcal E)+J\Lambda^{(\mathbf{w})}_\text{max}\phi(\langle V\rangle).
\end{equation*}
For simplicity, we will assume that mean-field solutions are stationary.
By setting the bias to $\mathcal{E}_c=\beta-J\Lambda^{(\mathbf{w})}_\text{max}\phi(\beta)$ where $\beta$ is an inflection point of the nonlinear function $\phi(x)$, we impose $\langle V\rangle=\beta$; this gives the first condition in Eq.~(\ref{eqn:CriticalityConditions}).
Varying $\mathcal{E}$ biases the firing rate of the spiking network, with $\mathcal E < \mathcal E_c$ priming the network activity for lower firing rates and $\mathcal E > \mathcal E_c$ inducing higher firing rates.
At $\langle V\rangle=\beta$, the second-derivative of $\phi(x)$ vanishes (i.e.~$\phi''(\beta)=0$).
Near this inflection point, the stationary values of the average membrane potential are given by
\begin{equation}
    \langle V\rangle=\beta\pm\sqrt{\frac{6(1-J\Lambda^{(\mathbf{w})}_\text{max}\phi'(\beta))}{J\Lambda^{(\mathbf{w})}_\text{max}\phi^{(3)}(\beta)}},\label{eqn:HT_mfV}
\end{equation}
where $\phi^{(n)}(\beta)$ is the $n^{\text{th}}$ derivative of $\phi(x)$ evaluated at $x=\beta$.
The condition $1-J\Lambda_\text{max}^{(\mathbf{w})}\phi'(\beta)=0$ defines the transition from a single stationary point to the bistable regime; this is the second condition in Eq.~(\ref{eqn:CriticalityConditions}).
Recalling that $\langle \dot{n}_i\rangle\approx\phi(\langle V_i\rangle)$, the average firing rate of the simulated spiking networks can be seen to recapitulate the predictions of the mean-field theory qualitatively (Fig.~\ref{fig:PhaseDiagram}B), which can be seen visually in raster plots of simulated spiking data (Fig.~\ref{fig:PhaseDiagram}C-F). 
Note that the stationary mean-field values for the response variables $\tilde{V}_i$, $\tilde{n}_i$, and $\tilde{x}_i$ are all 0.

After performing a second-order functional Taylor expansion of the action $S$ around the mean-field solutions, the Gaussian-approximated action $S_G$ can be rewritten compactly in block matrix notation as
\begin{subequations}
\begin{equation*}
    S_G[\mathbf{y}] = \frac{1}{2}\mathbf{y}^T\mathbf{A}\mathbf{y} = \frac{1}{2}\int dt~dt' \mathbf{y}^T(t) \mathbf{A}(t,t')\mathbf{y}(t')\label{subeqn:GPA_short_action}
\end{equation*}
\begin{align*}
    &\mathbf{y}(t)=\left[\mathbf{\tilde{V}}(t),\Delta\mathbf{V}(t),\mathbf{\tilde{n}}(t),\Delta\mathbf{\dot{n}}(t),\mathbf{\tilde{x}}(t),\Delta \mathbf{x}(t)\right]^T\nonumber\\
    &=\left[\mathbf{\tilde{V}}(t),\mathbf{V}(t)-\langle \mathbf{V}\rangle,\mathbf{\tilde{n}}(t),\mathbf{\dot{n}}(t)-\langle\mathbf{\dot{n}}\rangle,\mathbf{\tilde{x}}(t),\mathbf{x}(t)-\langle\mathbf{x}\rangle\right]^T.\label{subeqn:GPA_var_shorthand}
\end{align*}
\end{subequations}
We assume that the correlations for the noise process $\mathbf{x}$ are time-translation.

The desired correlations and response functions $\mathbf{\Delta}(t,t')$ (``propagator functions,'' more generally) are given by the inverse of the matrix function $\mathbf{A}(t,t')$, with the matrix inverse $\Delta(t,t')$ being defined through the relation
\be
    \int dt'' ~\mathbf{A}(t,t'')\mathbf{\Delta}(t'',t')= \delta(t-t')\mathbf{I}.\nonumber
\ee

The matrix $\mathbf{A}(t,t')$ in the Gaussian-approximated action is proportional to $\delta(t-t')$, reducing the inversion relation above into a simple matrix inversion.
The matrix inversion is easiest to handle in the frequency domain; we thus Fourier transform the matrix, invert it, and then take the inverse Fourier transform to get the expressions for the response functions and covariance functions given in Appendix \ref{app:rcf_GPA}.

\section{Response and covariance functions under the Gaussian process approximation}\label{app:rcf_GPA}

\begin{table}[t!]
    \begin{minipage}{\linewidth}
    \centering
        \begin{tabular}{|c|c|c || c|c|c |} 
            \hline
            Field & \multirow{2}{*}{Statistic} & Equation &Field & \multirow{2}{*}{Statistic} & Equation \\
            Variable & & Link & Variable & & Link \\
            [0.5ex] 
            \hline\hline
            \multirow{2}{*}{$\tilde{\Delta}_{\mathbf{V}\mathbf{V}}$} & \multirow{2}{*}{$\tilde{\mathbf{C}}^{\mathbf{V}\mathbf{V}}(\omega)$} & \multirow{2}{*}{\ref{eqn:APP_CVV_omega} } & \multirow{2}{*}{$\tilde{\Delta}_{\mathbf{V}\tilde{\mathbf{n}}}$} & \multirow{2}{*}{$\tilde{\mathbf{R}}^{\mathbf{V}\tilde{\mathbf{n}}}(\omega)$} & \multirow{2}{*}{\ref{eqn:APP_RVnt_omega}}\\
            &  & & &  &  \\
            \hline
            \multirow{2}{*}{$\tilde{\Delta}_{\mathbf{V}\dot{\mathbf{n}}}$} & \multirow{2}{*}{$\tilde{\mathbf{C}}^{\mathbf{V}\dot{\mathbf{n}}}(\omega)$} & \multirow{2}{*}{\ref{eqn:APP_CVn_omega}} & \multirow{2}{*}{$\tilde{\Delta}_{\tilde{\mathbf{n}}\mathbf{V}}$} & \multirow{2}{*}{$\tilde{\mathbf{R}}^{\tilde{\mathbf{n}}\mathbf{V}}(\omega)$} & \multirow{2}{*}{\ref{eqn:APP_RntV_omega}} \\
            &  & & &  &  \\
            \hline
            \multirow{2}{*}{$\tilde{\Delta}_{\mathbf{V}\mathbf{x}}$} & \multirow{2}{*}{$\tilde{\mathbf{C}}^{\mathbf{V}\mathbf{x}}(\omega)$} & \multirow{2}{*}{\ref{eqn:APP_CVx_omega}} & \multirow{2}{*}{$\tilde{\Delta}_{\mathbf{V}\tilde{\mathbf{x}}}$} & \multirow{2}{*}{$\tilde{\mathbf{R}}^{\mathbf{V}\tilde{\mathbf{x}}}(\omega)$} & \multirow{2}{*}{\ref{eqn:APP_RVxt_omega}}\\
            &  & & &  & \\
            \hline
            \multirow{2}{*}{$\tilde{\Delta}_{\dot{\mathbf{n}}\dot{\mathbf{n}}}$} & \multirow{2}{*}{$\tilde{\mathbf{C}}^{\dot{\mathbf{n}}\dot{\mathbf{n}}}(\omega)$}  & \multirow{2}{*}{\ref{eqn:APP_Cnn_omega}} & \multirow{2}{*}{$\tilde{\Delta}_{\tilde{\mathbf{
            x}}\mathbf{V}}$} & \multirow{2}{*}{$\tilde{\mathbf{R}}^{\tilde{\mathbf{x}}\mathbf{V}}(\omega)$} & \multirow{2}{*}{\ref{eqn:APP_RxtV_omega}}\\
            &  & & &  & \\
            \hline
            \multirow{2}{*}{$\tilde{\Delta}_{\dot{\mathbf{n}}\mathbf{x}}$} & \multirow{2}{*}{$\tilde{\mathbf{C}}^{\dot{\mathbf{n}}\mathbf{x}}(\omega)$} & \multirow{2}{*}{\ref{eqn:APP_Cnx_omega}} & \multirow{2}{*}{$\tilde{\Delta}_{\dot{\mathbf{n}}\tilde{\mathbf{n}}}$} & \multirow{2}{*}{$\tilde{\mathbf{R}}^{\dot{\mathbf{n}}\tilde{\mathbf{n}}}(\omega)$}  & \multirow{2}{*}{\ref{eqn:APP_Rnnt_omega}} \\
            &  & & &  & \\
            \hline
            \multirow{2}{*}{$\tilde{\Delta}_{\mathbf{x}\mathbf{x}}$} & \multirow{2}{*}{$\tilde{\mathbf{C}}^{\mathbf{x}\mathbf{x}}(\omega)$}  & \multirow{2}{*}{\ref{eqn:APP_Cxx_omega}} & \multirow{2}{*}{$\tilde{\Delta}_{\tilde{\mathbf{n}}\dot{\mathbf{n}}}$} & \multirow{2}{*}{$\tilde{\mathbf{R}}^{\tilde{\mathbf{n}}\dot{\mathbf{n}}}(\omega)$} & \multirow{2}{*}{\ref{eqn:APP_Rntn_omega}}\\
            &  & & &  & \\ 
            \hline
            \multirow{2}{*}{$\tilde{\Delta}_{\tilde{\mathbf{V}}\mathbf{V}}$} & \multirow{2}{*}{$\tilde{\mathbf{R}}^{\tilde{\mathbf{V}}\mathbf{V}}(\omega)$} &\multirow{2}{*}{\ref{eqn:APP_RVtV_omega}} & \multirow{2}{*}{$\tilde{\Delta}_{\dot{\mathbf{n}}\tilde{\mathbf{x}}}$} & \multirow{2}{*}{$\tilde{\mathbf{R}}^{\dot{\mathbf{n}}\tilde{\mathbf{x}}}(\omega)$} & \multirow{2}{*}{\ref{eqn:APP_Rnxt_omega}}\\
            &  & & &  & \\ 
            \hline
            \multirow{2}{*}{$\tilde{\Delta}_{\mathbf{V}\tilde{\mathbf{V}}}$} & \multirow{2}{*}{$\tilde{\mathbf{R}}^{\mathbf{V}\tilde{\mathbf{V}}}(\omega)$} & \multirow{2}{*}{\ref{eqn:APP_RVVt_omega}} & \multirow{2}{*}{$\tilde{\Delta}_{\tilde{\mathbf{x}}\dot{\mathbf{n}}}$} & \multirow{2}{*}{$\tilde{\mathbf{R}}^{\tilde{\mathbf{x}}\dot{\mathbf{n}}}(\omega)$} & \multirow{2}{*}{\ref{eqn:APP_Rxtn_omega}}\\
            &  & & &  & \\ 
            \hline
            \multirow{2}{*}{$\tilde{\Delta}_{\dot{\mathbf{n}}\tilde{\mathbf{V}}}$} & \multirow{2}{*}{$\tilde{\mathbf{R}}^{\dot{\mathbf{n}}\tilde{\mathbf{V}}}(\omega)$} & \multirow{2}{*}{\ref{eqn:APP_RnVt_omega}} & \multirow{2}{*}{$\tilde{\Delta}_{\mathbf{x}\tilde{\mathbf{x}}}$} & \multirow{2}{*}{$\tilde{\mathbf{R}}^{\mathbf{x}\tilde{\mathbf{x}}}(\omega)$}  & \multirow{2}{*}{\ref{eqn:APP_Rxxt_omega}}\\
            &  & & &  & \\
            \hline
            \multirow{2}{*}{$\tilde{\Delta}_{\tilde{\mathbf{V}}\dot{\mathbf{n}}}$} & \multirow{2}{*}{$\tilde{\mathbf{R}}^{\tilde{\mathbf{V}}\dot{\mathbf{n}}}(\omega)$} & \multirow{2}{*}{\ref{eqn:APP_RVtn_omega}} & \multirow{2}{*}{$\tilde{\Delta}_{\tilde{\mathbf{x}}\mathbf{x}}$} & \multirow{2}{*}{$\tilde{\mathbf{R}}^{\tilde{\mathbf{x}}\mathbf{x}}(\omega)$}  & \multirow{2}{*}{\ref{eqn:APP_Rxtx_omega}}\\
            &  & & &  & \\
            \hline
        \end{tabular}
    \caption{\textbf{Expressions for the second-order statistics in frequency space}\label{tbl:FrequencyForms} This table provides a list of the relevant second order statistics and links to their formulae.} 
    \end{minipage}
    
\end{table}

In this appendix, we summarize the results from the Gaussian process approximation described in Appendix \ref{app:PathIntGPA}, shifting from the opaque field notation for the second order statistics introduced therein towards the more intuitive notation used in Appendix \ref{app:rcf_GPA}. 
Note that the covariance functions $\tilde{\mathbf{C}}^{\mathbf{P}\mathbf{Q}}(\omega)$ can be written compactly as sums of products of the response functions $\tilde{\mathbf{C}}^{\mathbf{P}'\mathbf{Q}'}(\omega)$ when in the frequency domain; it can be shown using the convolution theorem that the products of response functions $\tilde{\mathbf{R}}^{\mathbf{P}'\mathbf{Q}'}(\omega)$ in $\tilde{\mathbf{C}}^{\mathbf{P}\mathbf{Q}}(\omega)$ become convolutions of the response functions $\mathbf{R}^{\mathbf{P}\mathbf{Q}}(\tau)$ in the time domain.

If the architecture of the spiking network and the input noise $\mathbf{x}$ share an eigenbasis, the expressions for the covariance and response functions can be represented as summations in eigenspace and the Fourier transform from the frequency domain $\omega$ to the time domain $\tau$ can be handled analytically. 
The requirement that both parts of the network share an eigenbasis amounts to requiring that 
\begin{equation}
    \label{eqn:SharedEigenbasis}
    \begin{split}
        \mathbf{J} = \mathbf{P}\mathbf{\Lambda}\mathbf{P}^T,\\
        \mathbf{w}^{x} = \mathbf{P}\mathbf{\Gamma}\mathbf{P}^T,\\
        \mathbf{P}\mathbf{P}^T = \mathbf{I},
    \end{split}
\end{equation}
where $\mathbf{\Lambda}$ and $\mathbf{\Gamma}$ are diagonal matrices holding the eigenvalues of their respective adjacency matrices and the matrix $\mathbf{P}$ of eigenvectors is the same for both systems.
In representing the response and covariance functions in eigenspace, it is convenient to define a set of eigentimescales corresponding to the evolution of network dynamics along its eigenmodes.
In particular, we have
\begin{align}
    \xi^{\alpha}_{\dot{\mathbf{n}}} := &\left(1-\lambda_\alpha^{(\mathbf{J})}\phi'(\langle V \rangle)\right)^{-1},\\
    \xi^{\alpha}_{\mathbf{x}} := & \left(r+3g\langle x \rangle^2-\lambda_\alpha^{(\mathbf{w})}\right)^{-1},
\end{align}
where $\lambda_\alpha^{(\mathbf{J})}$ and $\lambda_\alpha^{(\mathbf{w})}$ are the eigenvalues of $\mathbf{J}$ and $\mathbf{w}^x$, respectively, indexed by $\alpha \in 1, \dots, N$.

\begin{table}[t!]
    \begin{minipage}{\linewidth}        
        \centering
        \begin{tabular}{|c|c|c || c|c|c |} 
            \hline
            Field & \multirow{2}{*}{Statistic} &  Equation  &Field & \multirow{2}{*}{Statistic} & Equation \\
            Variable & & Link & Variable & & Link \\
            [0.5ex] 
            \hline\hline
            \multirow{2}{*}{$\Delta_{\mathbf{V}\mathbf{V}}$} & \multirow{2}{*}{$\mathbf{C}^{\mathbf{V}\mathbf{V}}(\tau)$} & \multirow{2}{*}{\ref{eqn:APP_CVV_eigen} } & \multirow{2}{*}{$\Delta_{\mathbf{V}\tilde{\mathbf{n}}}$} & \multirow{2}{*}{$\mathbf{R}^{\mathbf{V}\tilde{\mathbf{n}}}(\tau)$} & \multirow{2}{*}{\ref{eqn:APP_RVnt_eigen}} \\
            &  & & &  & \\
            \hline
            \multirow{2}{*}{$\Delta_{\mathbf{V}\dot{\mathbf{n}}}$} & \multirow{2}{*}{$\mathbf{C}^{\mathbf{V}\dot{\mathbf{n}}}(\tau)$} & \multirow{2}{*}{\ref{eqn:APP_CVn_eigen}} & \multirow{2}{*}{$\Delta_{\tilde{\mathbf{n}}\mathbf{V}}$} & \multirow{2}{*}{$\mathbf{R}^{\tilde{\mathbf{n}}\mathbf{V}}(\tau)$} & \multirow{2}{*}{\ref{eqn:APP_RntV_eigen}} \\
            &  & & &  & \\
            \hline
            \multirow{2}{*}{$\Delta_{\mathbf{V}\mathbf{x}}$} & \multirow{2}{*}{$\mathbf{C}^{\mathbf{V}\mathbf{x}}(\tau)$} & \multirow{2}{*}{\ref{eqn:APP_CVx_eigen}} & \multirow{2}{*}{$\Delta_{\mathbf{V}\tilde{\mathbf{x}}}$} & \multirow{2}{*}{$\mathbf{R}^{\mathbf{V}\tilde{\mathbf{x}}}(\tau)$} & \multirow{2}{*}{\ref{eqn:APP_RVxt_eigen}} \\
            &  & & &  &\\
            \hline
            \multirow{2}{*}{$\Delta_{\dot{\mathbf{n}}\dot{\mathbf{n}}}$} & \multirow{2}{*}{$\mathbf{C}^{\dot{\mathbf{n}}\dot{\mathbf{n}}}(\tau)$}  & \multirow{2}{*}{\ref{eqn:APP_Cnn_eigen}} & \multirow{2}{*}{$\Delta_{\tilde{\mathbf{
            x}}\mathbf{V}}$} & \multirow{2}{*}{$\mathbf{R}^{\tilde{\mathbf{x}}\mathbf{V}}(\tau)$} & \multirow{2}{*}{\ref{eqn:APP_RxtV_eigen}}\\
            &  & & &  &\\
            \hline
            \multirow{2}{*}{$\Delta_{\dot{\mathbf{n}}\mathbf{x}}$} & \multirow{2}{*}{$\mathbf{C}^{\dot{\mathbf{n}}\mathbf{x}}(\tau)$} & \multirow{2}{*}{\ref{eqn:APP_Cnx_eigen}} &\multirow{2}{*}{$\Delta_{\dot{\mathbf{n}}\tilde{\mathbf{n}}}$} & \multirow{2}{*}{$\mathbf{R}^{\dot{\mathbf{n}}\tilde{\mathbf{n}}}(\tau)$}  & \multirow{2}{*}{\ref{eqn:APP_Rnnt_eigen}} \\
            &  & & &  &\\
            \hline
            \multirow{2}{*}{$\Delta_{\mathbf{x}\mathbf{x}}$} & \multirow{2}{*}{$\mathbf{C}^{\mathbf{x}\mathbf{x}}(\tau)$}  & \multirow{2}{*}{\ref{eqn:APP_Cxx_eigen}} & \multirow{2}{*}{$\Delta_{\tilde{\mathbf{n}}\dot{\mathbf{n}}}$} & \multirow{2}{*}{$\mathbf{R}^{\tilde{\mathbf{n}}\dot{\mathbf{n}}}(\tau)$} & \multirow{2}{*}{\ref{eqn:APP_Rntn_eigen}}\\
            &  & & &  &\\ 
            \hline
            \multirow{2}{*}{$\Delta_{\tilde{\mathbf{V}}\mathbf{V}}$} & \multirow{2}{*}{$\mathbf{R}^{\tilde{\mathbf{V}}\mathbf{V}}(\tau)$} &\multirow{2}{*}{\ref{eqn:APP_RVtV_eigen}} & \multirow{2}{*}{$\Delta_{\dot{\mathbf{n}}\tilde{\mathbf{x}}}$} & \multirow{2}{*}{$\mathbf{R}^{\dot{\mathbf{n}}\tilde{\mathbf{x}}}(\tau)$} & \multirow{2}{*}{\ref{eqn:APP_Rnxt_eigen}}\\
            &  & & &  &\\ 
            \hline
            \multirow{2}{*}{$\Delta_{\mathbf{V}\tilde{\mathbf{V}}}$} & \multirow{2}{*}{$\mathbf{R}^{\mathbf{V}\tilde{\mathbf{V}}}(\tau)$} & \multirow{2}{*}{\ref{eqn:APP_RVVt_eigen}} & \multirow{2}{*}{$\Delta_{\tilde{\mathbf{x}}\dot{\mathbf{n}}}$} & \multirow{2}{*}{$\mathbf{R}^{\tilde{\mathbf{x}}\dot{\mathbf{n}}}(\tau)$} & \multirow{2}{*}{\ref{eqn:APP_Rxtn_eigen}}\\
            &  & & &  &\\ 
            \hline
            \multirow{2}{*}{$\Delta_{\dot{\mathbf{n}}\tilde{\mathbf{V}}}$} & \multirow{2}{*}{$\mathbf{R}^{\dot{\mathbf{n}}\tilde{\mathbf{V}}}(\tau)$} & \multirow{2}{*}{\ref{eqn:APP_RnVt_eigen}} & \multirow{2}{*}{$\Delta_{\mathbf{x}\tilde{\mathbf{x}}}$} & \multirow{2}{*}{$\mathbf{R}^{\mathbf{x}\tilde{\mathbf{x}}}(\tau)$}  & \multirow{2}{*}{\ref{eqn:APP_Rxxt_eigen}}\\
            &  & & &  &\\
            \hline
            \multirow{2}{*}{$\Delta_{\tilde{\mathbf{V}}\dot{\mathbf{n}}}$} & \multirow{2}{*}{$\mathbf{R}^{\tilde{\mathbf{V}}\dot{\mathbf{n}}}(\tau)$} & \multirow{2}{*}{\ref{eqn:APP_RVtn_eigen}} & \multirow{2}{*}{$\Delta_{\tilde{\mathbf{x}}\mathbf{x}}$} & \multirow{2}{*}{$\mathbf{R}^{\tilde{\mathbf{x}}\mathbf{x}}(\tau)$}  & \multirow{2}{*}{\ref{eqn:APP_Rxtx_eigen}}\\
            &  & & &  &\\
            \hline
        \end{tabular}\normalsize
        \caption{\textbf{Expressions for the second-order statistics in the time domain}\label{tbl:EigenForms} This table provides a list of the relevant second order statistics and links to their formulae.} 
    \end{minipage}
\end{table}

For the convenience of the reader, we provide lookup tables for the covariance and response statistics of the model.
Table \ref{tbl:FrequencyForms} provides a list of the covariance and response functions in frequency space using both notations for convenience, and links to their general formulae under the Gaussian process approximation; Table \ref{tbl:EigenForms} is a similar lookup table for the expressions in the time domain under the assumptions of Eq.~(\ref{eqn:SharedEigenbasis}).

\section{Mean-field approximation of the spike-spike covariances in the presence of latent input}

Ref.~\cite{brinkman_phase_2025} calculated the scaling forms for the decay of the mean firing rates using renormalization group (RG) techniques, which we use in this work because we expect the input from the latent network to average out over trials.
While this work did not estimate the scaling form for the covariance, in the absence of latent input the scaling form should follow from the expected scaling form for the Ising-like phase transition. 
However, calculating the covariance in the presence of input would require a more sophisticated RG calculation than presented in \cite{brinkman_phase_2025}, which is beyond the scope of this work.
However, because we focus on $d=3$, for which the critical exponents deviate from the mean-field predictions by small amounts, we estimate the scaling behavior of the covariance by calculating Gaussian fluctuation corrections to the mean-field approximation.
i.e., we approximate the stochastic processes for the spike trains and input network as Gaussian fluctuations around the mean-field estimates.
This can be systematically computed by representing the joint probability of the membrane potentials, spike trains, and input processes as a field theoretic path integral.
Details of this calculation are given in the Supplementary Information, and further information about the path integral formalism is available in Refs.~\cite{ocker_linking_2017,brinkman_phase_2025}.

The key result we highlight here is the population-averaged spike autocovariance, which can be expressed in terms of an integral over the eigenvalue density $\rho(\lambda)$ of the adjacency matrix $w_{ij}$: 
\begin{widetext}
\begin{align}
&C_{\dot{n}\dot{n}}(t) \nonumber \\
&= \phi(V_\infty)\delta(t) + \int_{-2d}^{2d} d\lambda~\rho(\lambda) \Bigg[ \frac{\phi'(V_\infty)\phi(V_\infty)}{2} \frac{J(\lambda-1)(2-J(\lambda-1)\phi'(V_\infty))}{1-J(\lambda-1)\phi'(V_\infty)}e^{-(1-J(\lambda-1) \phi'(V_\infty))|t|/\tau} \label{eqn:Cspkspkfull} \\
  & \hspace{3.0cm} + \frac{(\sigma \phi'(V_\infty))^2}{(r+3gx_\infty^2 - \lambda)^2 - (1 - J(\lambda-1)\phi'(V_\infty))^2}\left( \frac{e^{-(1-J(\lambda-1)\phi'(V_\infty))|t|/\tau}}{1-J(\lambda-1)\phi'(V_\infty)} - \frac{e^{-(r+3gx_\infty^2-\lambda)|t|/\tau}}{r+3gx_\infty^2-\lambda}\right) \Bigg],  \nonumber
\end{align}    
\end{widetext}
where $V_\infty$ is the steady-state mean of the membrane potential and $x_\infty$ is the steady state mean of the latent input (within the mean-field approximation).
As we focus on the case of subcritical or critical latent input, we set $x_\infty = 0$.
In writing Eq.~(\ref{eqn:Cspkspkfull}), we have exploited the fact that the spiking network and the input network are both arranged on the same $d$-dimensional lattice to express the covariance in terms of a single eigenvalue integral. 
In general these two networks would have different individual eigenspaces of their connectivity matrices, and the resulting expression would be much more complicated.

\begin{figure*}
\includegraphics[width=\linewidth]{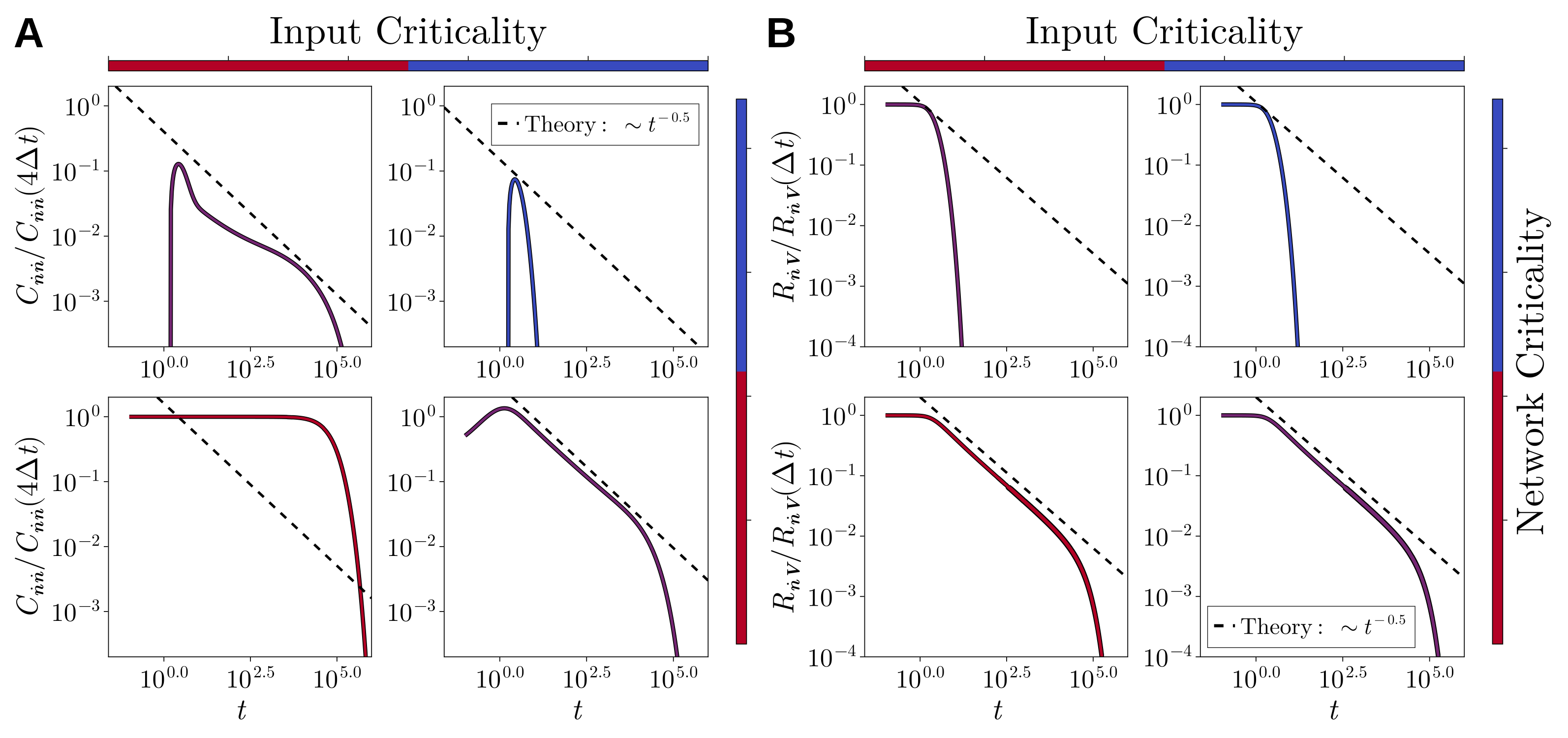}
\caption{\label{fig:SpkGridMF} \textbf{Mean-field predictions for the covariance and response functions}
Each color bar denotes whether the given subsystem is close to criticality (red) or far from criticality (blue).
(A) Spike-spike covariance functions corresponding to the combinations of the input and spiking network being near or far from their respective critical points.
(B) The differential response of the average spike rate in response to a perturbation of size $\Delta V = \pm10$ to the membrane potential of each neuron.
Spiking network: far-from-critical (blue) with $J_c -  J=0.5$, near-critical (red) with $J_c -  J = 2\times10^{-5}$.
Input network: far-from-critical (blue) with $r-r_c=1.0$, near-critical (red) with $r-r_c=10^{-5}$.}
\end{figure*}

In a similar manner to Fig.~\ref{fig:SpkGrid}, we compare the full covariance to the mean-field response (as calculated in \cite{brinkman_phase_2025}) in Fig.~\ref{fig:SpkGridMF}.
To calculate the covariances we use the full integral given in Eq.~(\ref{eqn:Cspkspkfull}).
Similar to simulations, the expected power-law tail of ($t^{-1/2}$ in $d=3$ for mean-field, $t^{-0.5105}$ in simulations) can be difficult to observe, showing what may appear to be a different power-law behavior being being cut off.
The mean-field calculation also predicts a plateau at times much less than ${\rm min}(\xi_{\rm spk},\xi_{\rm lat})$, whereas the simulations show a slight decaying power-law behavior, suggesting that fluctuations may renormalize this plateau. 
Using the exponent estimated in Fig.~\ref{fig:SpkGrid}, the renormalized behavior could be $t^{-2z_\ast \eta_\ast}$ or $t^{-4\eta_\ast}$. 
In $d=3$ lattices, $2 z_\ast \eta_\ast \approx 4 \eta_\ast \approx 0.14$, consistent with the simulation estimate of $t^{-0.13}$, while in mean-field theory $\eta_\ast = 0$, giving $t^{-0}$, consistent with the mean-field approximation.

\subsection*{Derivation of the combined scaling form}

We derive the general form of the mean-field scaling relation of the spike-spike covariance in the presence of input from the external population, Eq.~(\ref{eqn:Cnn-full_MFScaling}).
We begin from Eq.~(\ref{eqn:Cspkspkfull}), which estimates the covariance due to Gaussian fluctuations around the mean-field approximation of the network.
Because we expect scaling to hold only at long times, we ignore the delta function contribution at $t=0$ and focus on the remaining terms.

We first tackle the term
\begin{align*}
    &\int_{-2d}^{2d} d\lambda~\rho(\lambda) \Bigg[ \frac{\phi_1\phi_0}{2} \frac{J(\lambda-1)(2-J(\lambda-1)\phi_1)}{1-J(\lambda-1)\phi_1} \\
    & \hspace{5.0cm} \times e^{-(1-J(\lambda-1) \phi_1)|t|/\tau}\Bigg],
\end{align*}
where $\phi_n \equiv \phi^{(n)}(V_\infty)$ is a shorthand for derivatives of the nonlinearity evaluated at the steady-state membrane potential.
This term arises from the recurrent interactions from within the network, and depends on the state of the latent input process only through the mean input $x_\infty$.
In what follows we assume the latent input is not supercritical, such that $x_\infty = 0$.

To derive the scaling form, we change variables to $\gamma = 2d - \lambda$, which runs from $0$ to $4d$.
We define the coherence time by $\xi_{\rm spk}^{-1} = (1 - (2d-1) J \phi_1)/\tau$.
\begin{widetext}
\begin{align*}
&\int_{-2d}^{2d} d\lambda~\rho(\lambda) \Bigg[ \frac{\phi_1\phi_0}{2} \frac{J(\lambda-1)(2-J(\lambda-1)\phi_1)}{1-J(\lambda-1)\phi_1}e^{-(1-J(\lambda-1) \phi_1)|t|/\tau}\Bigg]\\
&= \frac{\phi_1\phi_0}{2} \int_{0}^{4d} d\gamma~\rho(2d-\gamma) \Bigg[ \frac{J(2d-1 - \gamma)(1+\tau\xi_{\rm spk}^{-1} + J\gamma\phi_1)}{\tau \xi_{\rm spk}^{-1} + J\gamma\phi_1}e^{-(\xi_{\rm spk}^{-1} + J\phi_1 \gamma/\tau)|t|}\Bigg]\\
&\sim \frac{\phi_0}{2} \left(\frac{1-\tau \xi^{-1}_{\rm spk}}{2d-1}\right) \int_{0}^{\infty} d\gamma~\gamma^{d/2-1} \Bigg[ \frac{(2d-1 - \gamma)\left(1+\tau \xi_{\rm spk}^{-1} +\gamma\left(\frac{1-\tau \xi_{\rm spk}^{-1}}{2d-1}\right)\right)}{\tau\xi_{\rm spk}^{-1} + \gamma \left(\frac{1-\tau\xi_{\rm spk}^{-1}}{2d-1}\right)}e^{-\left(\xi_{\rm spk}^{-1} + \frac{\gamma}{\tau} \left(\frac{1-\tau\xi_{\rm spk}^{-1}}{2d-1}\right)\right)|t|}\Bigg].
\end{align*}
We expressed terms involving $\phi_1$ in terms of the coherence time $\xi_{\rm spk}^{-1}$; in doing so, all factors of $J$ have also canceled out, leaving an expression purely in terms of the coherence time, which measure the distance to the network's intrinsic critical point.
The leading order scaling originates from the contribution of the integral near $\gamma = 0$, hence the replacement of $\rho(2d-\gamma) \simeq \gamma^{d/2-1}$ and the extension of the integration limits from $4d$ to $\infty$.
This integral can be evaluated in a series around $\xi^{-1}_{\rm spk} = 0$, but to obtain the general leading order scaling, which works when the integral cannot easily be evaluated, we make a change of variables $\gamma = \tau \xi_{\rm spk}^{-1}\hat{\gamma}$.
This gives
\begin{align*}
&\int_{-2d}^{2d} d\lambda~\rho(\lambda) \Bigg[ \frac{\phi_1\phi_0}{2} \frac{J(\lambda-1)(2-J(\lambda-1)\phi_1)}{1-J(\lambda-1)\phi_1}e^{-(1-J(\lambda-1) \phi_1)|t|/\tau}\Bigg]\\
&\sim \frac{\phi_0}{2} \left(\frac{1-\tau \xi^{-1}_{\rm spk}}{2d-1}\right) \xi_{\rm spk}^{-d/2+1}\int_{0}^{\infty} d\hat{\gamma}~\hat{\gamma}^{d/2-1} \Bigg[ \frac{(2d-1 - \tau \xi_{\rm spk}^{-1}\hat{\gamma})\left(1+\tau \xi_{\rm spk}^{-1} +\tau \xi_{\rm spk}^{-1}\hat{\gamma}\left(\frac{1-\xi_{\rm spk}^{-1}}{2d-1}\right)\right)}{1 + \hat{\gamma} \left(\frac{1-\tau\xi_{\rm spk}^{-1}}{2d-1}\right)}e^{-\left(1 + \hat{\gamma} \left(\frac{1-\tau\xi_{\rm spk}^{-1}}{2d-1}\right)\right)|t|/\xi_{\rm spk}}\Bigg]\\
&\simeq \xi_{\rm spk}^{-d/2+1}\int_{0}^{\infty} d\hat{\gamma}~\hat{\gamma}^{d/2-1} \Bigg[ \frac{2d-1}{1 +  \frac{\hat{\gamma}}{2d-1}}e^{-\left(1 +  \frac{\hat{\gamma}}{2d-1}\right)|t|/\xi_{\rm spk}}\Bigg]\\
&= \xi_{\rm spk}^{-d/2+1} F(t/\xi_{\rm spk}),
\end{align*}
where in the second-to-last line we assume $|t|/\xi_{\rm spk}$ is order $1$ and neglect the remaining dependence on the coherence time inside the integral.
We can define $F(t/\xi_{\rm spk}) = (t/\xi_{\rm spk})^{-d/2+1}G(t/\xi_{\rm spk})$ to write this scaling form in the form $t^{-d/2+1} F(t/\xi_{\rm spk})$, which indicates the power-law decay of the covariance at long times, which is cut off if the coherence time is finite.

The next contribution to the covariance is more involved, depending on both the distance of the spiking network to its critical point and the distance of the latent input process to its critical point:
\begin{align*}
& \int_{-2d}^{2d} d\lambda~\rho(\lambda) \frac{(\sigma \phi_1)^2}{(r - \lambda)^2 - (1 - J(\lambda-1)\phi_1)^2}\left( \frac{e^{-(1-J(\lambda-1)\phi_1))|t|/\tau}}{1-J(\lambda-1)\phi_1} - \frac{e^{-(r-\lambda)|t|/\tau}}{r-\lambda}\right) \Bigg].  
\end{align*} 
As stated above, we have assumed explicitly that $x_\infty = 0$, which corresponds to $r - 2d \geq 0$.
Because this is positive, we may define the coherence time of the latent input process by $\xi_{\rm lat}^{-1} \equiv (r - 2d)/\tau$.

As in the previous term, we proceed by changing variables to $\gamma = 2d - \lambda$.
Let us first express the prefactor of the exponential terms in terms of the two coherence lengths.
The denominator is
\begin{align*}
&(\tau\xi_{\rm lat}^{-1} + \gamma)^2 - ( \tau \xi_{\rm spk}^{-1} + J\phi_1 \gamma)^2\\
&= (\tau\xi_{\rm lat})^{-2} - (\tau\xi_{\rm spk})^{-2} + 2 \gamma(\tau\xi_{\rm lat}^{-1} - \tau\xi_{\rm spk}^{-1}J\phi_1) + \gamma^2(1 - J^2\phi_1^2)\\
&= (\tau\xi_{\rm lat})^{-2} - (\tau\xi_{\rm spk})^{-2} + 2 \gamma\left(\tau\xi_{\rm lat}^{-1} - \tau\xi_{\rm spk}^{-1} \left(\frac{1-\tau \xi_{\rm spk}^{-1}}{2d-1}\right)\right) + \gamma^2\left(1 - \left(\frac{1-\tau \xi_{\rm spk}^{-1}}{2d-1}\right)^2\right).
\end{align*}
As in the previous term, we will change variables to $\gamma = \tau\xi_{\rm spk}^{-1} \hat{\gamma}$, and we will define the ratio $\kappa \equiv \xi_{\rm lat}^{-1}/\xi_{\rm spk}^{-1} = \xi_{\rm spk}/\xi_{\rm lat}$.
The denominator may then be written as
\begin{align*}
   &\tau^2 \xi_{\rm spk}^{-2}\left( \kappa^{2} - 1 + 2 \hat{\gamma}\left(\kappa -  \left(\frac{1-\tau \xi_{\rm spk}^{-1}}{2d-1}\right)\right) + \hat{\gamma}^2\left(1 - \left(\frac{1-\tau \xi_{\rm spk}^{-1}}{2d-1}\right)^2\right)\right)\\
   &\simeq \tau^2\xi_{\rm spk}^{-2}\left( \kappa^{2} - 1 + 2 \hat{\gamma}\left(\kappa -  \frac{1}{2d-1}\right) + \hat{\gamma}^2\left(1 - \left(\frac{1}{2d-1}\right)^2\right)\right)
\end{align*}
where in the last line we assume that $\kappa$ could be order $1$, and we neglect the remaining $\xi^{-1}_{\rm spk}$ dependence, assuming the spiking network is sufficiently close to its intrinsic critical point.

We now tackle the difference of exponentials:
\begin{align*}
    &\frac{e^{-(1-J(\lambda-1)\phi_1)|t|/\tau}}{1-J(\lambda-1)\phi_1} - \frac{e^{-(r-\lambda)|t|}}{r-\lambda}\\
&= \frac{e^{-\left(\xi_{\rm spk}^{-1} + \left(\frac{1-\tau \xi_{\rm spk}^{-1}}{2d-1}\right)\gamma\right)|t|}}{\tau\xi_{\rm spk}^{-1} + \left(\frac{1-\tau \xi_{\rm spk}^{-1}}{2d-1}\right)\gamma} - \frac{e^{-(\xi_{\rm lat}^{-1} + \gamma)|t|}}{\tau\xi_{\rm lat}^{-1} + \gamma}\\
&= \frac{e^{-\left(1 + \left(\frac{1-\tau \xi_{\rm spk}^{-1}}{2d-1}\right)\hat{\gamma}\right)|t|/\xi_{\rm spk}}}{\tau\xi_{\rm spk}^{-1}\left(1 + \left(\frac{1-\tau \xi_{\rm spk}^{-1}}{2d-1}\right)\hat{\gamma}\right)} - \frac{e^{-(\kappa + \hat{\gamma})|t|/\xi_{\rm spk}}}{\tau\xi_{\rm spk}^{-1}\left(\kappa + \hat{\gamma}\right)}\\
& \simeq \frac{1}{\tau \xi_{\rm spk}^{-1}}\left[\frac{e^{-\left(1 + \frac{\hat{\gamma}}{2d-1}\right)|t|/\xi_{\rm spk}}}{1 + \frac{\hat{\gamma}}{2d-1}} - \frac{e^{-(\kappa + \hat{\gamma})|t|/\xi_{\rm spk}}}{\kappa + \hat{\gamma}}\right]
\end{align*}
Putting everything together gives
\begin{align*}
& \xi_{\rm spk}^{-d/2+3}\int_{0}^{\infty} d\hat{\gamma}~\hat{\gamma}^{d/2-1} \frac{\left(\frac{\sigma }{J(2d-1)}\right)^2}{\tau^3\left( \kappa^{2} - 1 + 2 \hat{\gamma}\left(\kappa -  \frac{1}{2d-1}\right) + \hat{\gamma}^2\left(1 - \left(\frac{1}{2d-1}\right)^2\right)\right)}\left[\frac{e^{-\left(1 + \frac{\hat{\gamma}}{2d-1}\right)|t|/\xi_{\rm spk}}}{1 + \frac{\hat{\gamma}}{2d-1}} - \frac{e^{-(\kappa + \hat{\gamma})|t|/\xi_{\rm spk}}}{\kappa + \hat{\gamma}}\right] \Bigg].  
\end{align*} 
The prefactor of $\xi_{\rm spk}^{-d/2+3}$ makes it appear that this scaling form is divergent as $\xi_{\rm spk} \rightarrow \infty$ in the $d = 3$ case that we consider in this work. 
However, the dependence of the scaling form on the ratio $\kappa \equiv \xi_{\rm spk}/\xi_{\rm lat}$ cannot be ignored when considering the asymptotic limit of the scaling function. 
In the case $\kappa \gg 1$, we may keep only the $\kappa^2$ term in the denominator and neglect the exponential term $e^{-(\kappa + \hat{\gamma})|t|/\xi_{\rm spk}}/(\kappa+\hat{\gamma})$.
In the case $\kappa \ll 1$ we set $\kappa = 0$, in which case the term $\exp(-\hat{\gamma}|t|/\xi_{\rm spk})/\hat{\gamma}$ will dominant the integral, motivating another change of variables to $y = \hat{\gamma} |t|/\xi_{\rm spk}$.
The $\kappa = 1$ case is similar to the $\kappa \ll 1$ case because of the cancellation of $\kappa^2-1$ in the denominator, leading to $\hat{\gamma} = 0$ to dominate the integral and allowing a similar change of variables to $y = \hat{\gamma} |t|/\xi_{\rm spk}$.
We can summarize all three limits by a scaling form
$$t^{-d/2+1} {\rm min}(\xi_{\rm spk},\xi_{\rm lat})^2 \mathcal G(t/\xi_{\rm spk})$$
which exhibits the same decay with time as the intrinsic spike-spike covariance.
However, two remarks are in order: first, the actual specific scaling function $\mathcal G$ is different in each of the three limits.
Notably, for $\kappa = 1$ it is $e^{-|t|/\xi_{\rm spk}}$ to leading order, while for $\kappa \ll 1$ the scaling function is constant.
The second remark is that in all cases the result is proportional to the square of a coherence time.
This means that in the mean-field approximation we do not get a finite result for the covariance in the double-critical limit $\xi_{\rm spk},~\xi_{\rm lat} \rightarrow \infty$.
Our simulation results suggest that the covariance is well-behaved in this regime, though this could be a symptom of finite-size effects. 
To resolve the question of even the qualitative behavior of the doubly-critical limit, methods beyond the mean-field approximation are necessary.

\end{widetext}

\section{Scaling of the variance to mean ratio with network size and time-bin size}

To determine the location of the critical point we estimate a local peak in the ratio of the variance of the spiking activity to the mean activity in the steady-state regime.
Because averages and variances can be estimated across population, time, and trials, estimates of these quantities from data may differ from analytic estimates by factors of network size or time-bin size.
Here we estimate the expected scaling for the specific choice of mean and variance estimators we use in this work.

First, we define the population mean of the stochastic spike trains on a single trial as
\begin{align*}
    \hat{\nu}(t) &= N^{-1}\sum_i \dot{n}_i(t).
\end{align*}
We then estimate the variance of the network activity as the variance of this quantity over a time window $[t_0,t_0+T]$:
\begin{align*}
\hat{\sigma}^2 &= T^{-1}\int_{t_0}^{t_0+T} {dt}~\hat{\nu}(t)^2 - T^{-2} \left(\int_{t_0}^{t_0+T} dt~\hat{\nu}(t)\right)^2.
\end{align*}
We now compute the trial-averages as expectations over the distribution of network activity.
We denote $\mathbb{E}[\dot{n}_i(t)] = \nu$ and $C_{ij}(t-s) = {\rm cov}(\dot{n}_i(t),\dot{n}_j(s)) = \mathbb{E}[\dot{n}_i(t)\dot{n}_j(s)] - \nu^2$, where $\nu$ is the steady-state mean, which is independent of time and neuron index, and $C_{ij}(t-s)$ is the cross-covariance of neurons $i$ and $j$ at times $t$ and $s$.
The expectation of the population mean is straightforward,
\begin{align*}
\mathbb{E}\left[T^{-1}\int_{t_0}^{t_0+T}dt~\hat{\nu}(t)\right] &= \nu.
\end{align*}

The expectation of the variance is more involved:
\begin{align*}
&\mathbb{E}[\hat{\sigma}^2]\\ &= T^{-1} \int dt~ \mathbb{E}[\hat{\nu}(t)^2] - T^{-2} \int dt ds~ \mathbb{E}[\hat{\nu}(t) \hat{\nu}(s)]\\
&= T^{-1} \int dt~ \mathbb{E}[\hat{\nu}(t)^2] - T^{-2} \int dt ds~ \mathbb{E}[\hat{\nu}(t) \hat{\nu}(s)]\\
&= T^{-1} N^{-2} \sum_{ij} \int dt~ \mathbb{E}[\dot{n}_i(t) \dot{n}_j(t)] \\
&~~~~- T^{-2}N^{-2} \sum_{ij} \int dt ds \mathbb{E}[\dot{n}_i(t) \dot{n}_j(s)]\\
&= T^{-1} N^{-2} \sum_{ij} \int dt~ \left(C_{ij}(0) + \nu^2\right) \\
& ~~~~ - T^{-2}N^{-2} \sum_{ij} \int dt ds~ \left(C_{ij}(t-s) + \nu^2\right)\\
&= \frac{1}{N^2}\sum_{ij}\Big[C_{ij}(0) \\
& ~~~~~~~~ - \frac{1}{T^2}\int_{t_0}^{t_0+T}dt \int_{t_0}^{t_0+T}ds~C_{ij}(t-s) \Big]
\end{align*}
Because our goal is only to estimate the scaling of the variance-to-mean ratio with population size $N$, duration $T$, and time-bin size $\Delta t_{\rm bin}$, we use only the leading order behavior of the covariance, which is $C_{ij}(t-s) \approx \nu \delta_{ij} \delta(t-s)$.
We see that the first term will give a factor of $\delta(0)$, which we regularize by using the fact that the simulation data is discretized in bins of size $\Delta t_{\rm bin}$.
This regularization gives $\delta(0) \rightarrow 1/\Delta t_{\rm bin}$.
With this replacement we can evaluate the time integrals and sums to obtain
\begin{align*}
\mathbb{E}[\hat{\sigma}^2] &= \frac{\nu}{N}\left[\frac{1}{\Delta t_{\rm bin}} - \frac{1}{T}\right]\\
&\approx \frac{\nu}{N \Delta t_{\rm bin}},
\end{align*}
using the fact that $\Delta t_{\rm bin} \ll T$.

Finally, the ratio of the variance to the mean is
\begin{align*}
\frac{\mathbb{E}[\hat{\sigma}^2]}{\mathbb{E}[T^{-1} \int dt~\hat{\nu} ]} &\approx \frac{1}{N \Delta t_{\rm bin}},
\end{align*}
which shows that the variances given in Fig.~\ref{fig:HT_RidgeFitting} can be multiplied by $N\delta t_{\rm bin}$ to obtain estimates of the direct ratios of $C_{ii}(0)/\mathbb{E}[\dot{n}_i(t)]$.

\section{Response and covariance functions}

This Appendix lists the covariance and response statistics of the model under the saddle point approximation in Appendix \ref{app:PathIntGPA}.

\begin{widetext}

\begin{align}
    \label{eqn:APP_CVV_omega}
    \tilde{\mathbf{C}}^{\mathbf{V}\mathbf{V}}(\omega) = \tilde{\mathbf{R}}^{\mathbf{V}\tilde{\mathbf{n}}}(\omega)\,{\rm diag}(\phi(\langle V \rangle))\,\tilde{\mathbf{R}}^{\tilde{\mathbf{n}}\mathbf{V}}(\omega)+\tilde{\mathbf{R}}^{\mathbf{V}\tilde{\mathbf{V}}}(\omega)\tilde{\mathbf{C}}^{\mathbf{x}\mathbf{x}}(\omega)\tilde{\mathbf{R}}^{\tilde{\mathbf{V}}\mathbf{V}}(\omega)
\end{align}
\begin{align}
    \label{eqn:APP_CVn_omega}
    \tilde{\mathbf{C}}^{\mathbf{V}\dot{\mathbf{n}}}(\omega)=\tilde{\mathbf{R}}^{\dot{\mathbf{n}}\tilde{\mathbf{n}}}(\omega)\,{\rm diag}(\phi(\langle V \rangle))\, \tilde{\mathbf{R}}^{\tilde{\mathbf{n}}\mathbf{V}}(\omega)+\tilde{\mathbf{R}}^{\dot{\mathbf{n}}\tilde{\mathbf{V}}}(\omega)\tilde{\mathbf{C}}^{\mathbf{x}\mathbf{x}}(\omega)\tilde{\mathbf{R}}^{\tilde{\mathbf{V}}\mathbf{V}}(\omega)
\end{align}
\begin{align}
    \label{eqn:APP_CVx_omega}
    \tilde{\mathbf{C}}^{\mathbf{V}\mathbf{x}}(\omega)=-\tilde{\mathbf{C}}^{\mathbf{x}\mathbf{x}}(\omega)\tilde{\mathbf{R}}^{\tilde{\mathbf{V}}\mathbf{V}}(\omega)
\end{align}
\begin{align}
    \label{eqn:APP_Cnn_omega}
    \tilde{\mathbf{C}}^{\dot{\mathbf{n}}\dot{\mathbf{n}}}(\omega)=\tilde{\mathbf{R}}^{\dot{\mathbf{n}}\tilde{\mathbf{n}}}(\omega)\,{\rm diag}(\phi(\langle V \rangle))\, \tilde{\mathbf{R}}^{\tilde{\mathbf{n}}\dot{\mathbf{n}}}(\omega)+\tilde{\mathbf{R}}^{\dot{\mathbf{n}}\tilde{\mathbf{V}}}(\omega) \tilde{\mathbf{C}}^{\mathbf{x}\mathbf{x}}(\omega)\tilde{\mathbf{R}}^{\tilde{\mathbf{V}}\dot{\mathbf{n}}}(\omega)
\end{align}
\begin{align}
    \label{eqn:APP_Cnx_omega}
     \tilde{\mathbf{C}}^{\dot{\mathbf{n}}\mathbf{x}}(\omega)=-\tilde{\mathbf{C}}^{\mathbf{x}\mathbf{x}}(\omega)\tilde{\mathbf{R}}^{\tilde{\mathbf{V}}\dot{\mathbf{n}}}(\omega)
\end{align}
\begin{align}
    \label{eqn:APP_Cxx_omega}
    \tilde{\mathbf{C}}^{\mathbf{x}\mathbf{x}}(\omega) = 2\sigma^2\tilde{\mathbf{R}}^{\mathbf{x}\tilde{\mathbf{x}}}(\omega)\tilde{\mathbf{R}}^{\tilde{\mathbf{x}}\mathbf{x}}(\omega)
\end{align}
\begin{align}
    \label{eqn:APP_RVtV_omega}
    \tilde{\mathbf{R}}^{\tilde{\mathbf{V}}\mathbf{V}}(\omega)=\left[(i\omega+1)\mathbf{I}-{\rm diag}(\phi'(\langle V \rangle))\,\mathbf{J}\right]^{-1}
\end{align}
\begin{align}
    \label{eqn:APP_RVVt_omega}
    \tilde{\mathbf{R}}^{\mathbf{V}\tilde{\mathbf{V}}}(\omega)=\left[(-i\omega+1)\mathbf{I}-\mathbf{J}^{T}{\rm diag}(\phi'(\langle V \rangle))\right]^{-1}
\end{align}
\begin{align}
    \label{eqn:APP_RnVt_omega}
    \tilde{\mathbf{R}}^{\dot{\mathbf{n}}\tilde{\mathbf{V}}}(\omega)=\left[ (-i\omega+1)\mathbf{I}-{\rm diag}(\phi'(\langle V \rangle))\,\mathbf{J}^{T}\right]^{-1}{\rm diag}(\phi'(\langle V \rangle))
\end{align}
\begin{align}
    \label{eqn:APP_RVtn_omega}
    \tilde{\mathbf{R}}^{\tilde{\mathbf{V}}\dot{\mathbf{n}}}(\omega)=\left[(i\omega+1)\mathbf{I}-\mathbf{J}\,{\rm diag}(\phi'(\langle V \rangle))\right]^{-1}\mathbf{J}
\end{align}
\begin{align}
    \label{eqn:APP_RVnt_omega}
    \tilde{\mathbf{R}}^{\mathbf{V}\tilde{\mathbf{n}}}(\omega)=\mathbf{J}^{T}\left[(-i\omega+1)\mathbf{I}-{\rm diag}(\phi'(\langle V \rangle))\,\mathbf{J}^{T}\right]^{-1}
\end{align}
\begin{align}
    \label{eqn:APP_RntV_omega}
    \tilde{\mathbf{R}}^{\tilde{\mathbf{n}}\mathbf{V}}(\omega)=\left[(i\omega+1)\mathbf{I}-\mathbf{J}\,{\rm diag}(\phi'(\langle V \rangle))\right]^{-1}\mathbf{J}
\end{align}
\begin{align}
    \label{eqn:APP_RVxt_omega}
    \tilde{\mathbf{R}}^{\mathbf{V}\tilde{\mathbf{x}}}(\omega)=&{\rm diag}(1/\phi'(\langle V \rangle))\left[(-i\omega+1)\mathbf{I}-{\rm diag}(\phi'(\langle V \rangle))\,\mathbf{J}^{T}\right]^{-1}\nonumber\\
    &~~\times{\rm diag}(\phi'(\langle V \rangle))\left[ -\mathbf{w}^{x}+{\rm diag}(i\omega+r+3g\left(x_i^{\rm mf}\right)^2)\right]^{-1}
\end{align}
\begin{align}
    \label{eqn:APP_RxtV_omega}
    \tilde{\mathbf{R}}^{\tilde{\mathbf{x}}\mathbf{V}}(\omega)=&\left[\mathbf{w}^{x} + {\rm diag}(-i\omega+r+3g\left(x_i^{\rm mf}\right)^2)\right]^{-1}{\rm diag}(\phi'(\langle V \rangle))\nonumber\\
    &~~\times\left[(i\omega+1)\mathbf{I}-\mathbf{J}\,{\rm diag}(\phi'(\langle V \rangle))\right]^{-1}{\rm diag}(1/\phi'(\langle V \rangle))
\end{align}
\begin{align}
    \label{eqn:APP_Rnnt_omega}
    \tilde{\mathbf{R}}^{\dot{\mathbf{n}}\tilde{\mathbf{n}}}(\omega)=(-i\omega+1)\left[(-i\omega+1)\mathbf{I}-{\rm diag}(\phi'(\langle V \rangle))\,\mathbf{J}^{T}\right]^{-1}
\end{align}
\begin{align}
    \label{eqn:APP_Rntn_omega}
     \tilde{\mathbf{R}}^{\tilde{\mathbf{n}}\dot{\mathbf{n}}}(\omega)=(i\omega+1) \left[(i\omega+1)\mathbf{I}-\mathbf{J}\,{\rm diag}(\phi'(\langle V \rangle))\right]^{-1}
\end{align}
\begin{align}
    \label{eqn:APP_Rnxt_omega}
    \tilde{\mathbf{R}}^{\dot{\mathbf{n}}\tilde{\mathbf{x}}}(\omega)=&\left[ (-i\omega+1)\mathbf{I}-{\rm diag}(\phi'(\langle V \rangle))\,\mathbf{J}^{T}\right]^{-1}{\rm diag}(\phi'(\langle V \rangle))\nonumber\\
    &~~\times\left[ -\mathbf{w}^{x}+{\rm diag}(i\omega+r+3g\left(x_i^{\rm mf}\right)^2)\right]^{-1}
\end{align}
\begin{align}
    \label{eqn:APP_Rxtn_omega}
    \tilde{\mathbf{R}}^{\tilde{\mathbf{x}}\dot{\mathbf{n}}}(\omega)=&\left[ -\mathbf{w}^{x}+{\rm diag}(-i\omega+r+3g\left(x_i^{\rm mf}\right)^2)\right]^{-1}\nonumber\\
    &~~\times{\rm diag}(\phi'(\langle V \rangle))\left[(i\omega+1)\mathbf{I}-\mathbf{J}\,{\rm diag}(\phi'(\langle V \rangle))\right]^{-1}
\end{align}
\begin{align}
    \label{eqn:APP_Rxxt_omega}
    \tilde{\mathbf{R}}^{\mathbf{x}\tilde{\mathbf{x}}}(\omega)=[ -\mathbf{w}^{x}+{\rm diag}(i\omega+r+3g\left(x_i^{\rm mf}\right)^2)]^{-1}
\end{align}
\begin{align}
    \label{eqn:APP_Rxtx_omega}
    \tilde{\mathbf{R}}^{\tilde{\mathbf{x}}\mathbf{x}}(\omega)=[ - \mathbf{w}^{x} + {\rm diag}(-i\omega+r+3g\left(x_i^{\rm mf}\right)^2)]^{-1}
\end{align}

\begin{align}
    \label{eqn:APP_CVV_eigen}
        \mathbf{C}^{\mathbf{V}\mathbf{V}}_{ij}(\tau) = \sum_{\alpha} & \frac{{P}_{i\alpha}{P}_{j\alpha}}{2} \Bigg[\lambda_\alpha^2\xi^{\alpha}_{\dot{\mathbf{n}}}\phi(\langle V \rangle)e^{- |\tau|/\xi^{\alpha}_{\dot{\mathbf{n}}}} \nonumber\\
        &~~~+\frac{\sigma^2}{\left( \xi^{\alpha}_{\mathbf{x}}\right)^{-2} - \left(\xi^{\alpha}_{\dot{\mathbf{n}}}\right)^{-2}}\Bigg(  \xi^{\alpha}_{\dot{\mathbf{n}}}e^{ -|\tau|/\xi^{\alpha}_{\mathbf{\dot{n}}}}  - \xi^{\alpha}_{\mathbf{x}} e^{ -|\tau|/\xi^{\alpha}_{\mathbf{x}}}\Bigg)\Bigg]
\end{align}

\begin{align}
    \label{eqn:APP_CVn_eigen}
     {C}^{\mathbf{V}\dot{\mathbf{n}}}_{ij}(\tau) = \sum_{\alpha} &\frac{ {P}_{i\alpha} {P}_{j\alpha}}{2} \Bigg[ \lambda_\alpha\phi(\langle V \rangle)\left(\xi^{\alpha}_{\dot{\mathbf{n}}} + \text{sgn}(\tau)\right) e^{- |\tau|/\xi^{\alpha}_{\dot{\mathbf{n}}}} \nonumber\\
     &~~~+\frac{\sigma^2\phi'(\langle V \rangle)}{\left( \xi^{\alpha}_{\mathbf{x}}\right)^{-2} - \left(\xi^{\alpha}_{\dot{\mathbf{n}}}\right)^{-2}}\left(\xi^{\alpha}_{\dot{\mathbf{n}}}e^{ -|\tau|/\xi^{\alpha}_{\mathbf{\dot{n}}}} -\xi^{\alpha}_{\mathbf{x}} e^{ -|\tau|/\xi^{\alpha}_{\mathbf{x}}}  \right)\Bigg]
\end{align}

\begin{align}
    \label{eqn:APP_CVx_eigen}
         {C}^{\mathbf{V}\mathbf{x}}_{ij}(\tau) = \sum_{\alpha}& \frac{ {P}_{i\alpha} {P}_{j\alpha}}{2} \Bigg[\frac{2\sigma^2\xi^{\alpha}_{\mathbf{x}}}{\left(\xi^{\alpha}_{\dot{\mathbf{n}}}\right)^{-2}-\left( \xi^{\alpha}_{\mathbf{x}}\right)^{-2}  }\Bigg(\left( \frac{1}{\xi^{\alpha}_{\mathbf{x}}}-\frac{1}{\xi^{\alpha}_{\dot{\mathbf{n}}}}\right) e^{ -|\tau|/\xi^{\alpha}_{\mathbf{x}}}+\frac{2\Theta\big(\tau \big)}{\xi^{\alpha}_{\mathbf{x}}} e^{ -|\tau|/\xi^{\alpha}_{\mathbf{\dot{n}}}} \Bigg)\Bigg]
\end{align}

\begin{align}
    \label{eqn:APP_Cnn_eigen}
        {C}^{\dot{\mathbf{n}}\dot{\mathbf{n}}}_{ij}(\tau) = \sum_{\alpha} &{P}_{i\alpha}{P}_{j\alpha} \Bigg[\delta(\tau)+ \frac{\lambda_\alpha\phi(\langle V \rangle)\phi'(\langle V \rangle)(2-\lambda_\alpha\phi'(\langle V \rangle))}{2} \xi^{\alpha}_{\dot{\mathbf{n}}}e^{- |\tau|/\xi^{\alpha}_{\dot{\mathbf{n}}}} \nonumber\\
        &~~~+\frac{\left(\sigma\phi'(\langle V \rangle)\right)^2}{\left( \xi^{\alpha}_{\mathbf{x}}\right)^{-2} - \left(\xi^{\alpha}_{\dot{\mathbf{n}}}\right)^{-2}}\Bigg( \xi^{\alpha}_{\dot{\mathbf{n}}} e^{-|\tau|/\xi^{\alpha}_{\dot{\mathbf{n}}} }-\xi^{\alpha}_{\mathbf{x}} e^{ -|\tau|/\xi^{\alpha}_{\mathbf{x}}}  \Bigg)\Bigg]
\end{align}

\begin{align}
    \label{eqn:APP_Cnx_eigen}
       {C}^{\dot{\mathbf{n}}\mathbf{x}}_{ij}(\tau) = \sum_{\alpha} &\frac{{P}_{i\alpha}{P}_{j\alpha}}{2} \Bigg[\frac{\sigma^2\xi^{\alpha}_{\mathbf{x}}}{\left(\xi^{\alpha}_{\dot{\mathbf{n}}}\right)^{-2}-\left( \xi^{\alpha}_{\mathbf{x}}\right)^{-2}  }\Bigg(\left( \frac{1}{\xi^{\alpha}_{\mathbf{x}}}-\frac{1}{\xi^{\alpha}_{\dot{\mathbf{n}}}}\right) e^{ -|\tau|/\xi^{\alpha}_{\mathbf{x}}} +\frac{2\Theta(\tau)}{\xi^{\alpha}_{\mathbf{x}}} e^{ -|\tau|/\xi^{\alpha}_{\mathbf{\dot{n}}}} \Bigg)\Bigg]
\end{align}

\begin{align}
    \label{eqn:APP_Cxx_eigen}
        {C}^{\mathbf{x}\mathbf{x}}_{ij}(\tau) = \sum_{\alpha}& {P}_{i\alpha}{P}_{j\alpha}  \Bigg[\sigma^2\xi^{\alpha}_{\mathbf{x}} e^{ -|\tau|/\xi^{\alpha}_{\mathbf{x}}}\Bigg]
\end{align}

\begin{align}
    \label{eqn:APP_RVtV_eigen}
    {R}^{\tilde{\mathbf{V}}\mathbf{V}}_{ij}(\tau)=\sum_{\alpha} {P}_{i\alpha}{P}_{j\alpha}\Theta\big(\tau \big) e^{-|\tau|/\xi^{\alpha}_{\dot{\mathbf{n}}}}
\end{align}
\begin{align}
    \label{eqn:APP_RVVt_eigen}
    {R}^{\mathbf{V}\tilde{\mathbf{V}}}_{ij}(\tau) = \sum_{\alpha} {P}_{i\alpha}{P}_{j\alpha}\Theta\big(-\tau \big) e^{-|\tau|/\xi^{\alpha}_{\dot{\mathbf{n}}}}
\end{align}
\begin{align}
    \label{eqn:APP_RnVt_eigen}
    {R}^{\dot{\mathbf{n}}\tilde{\mathbf{V}}}_{ij}(\tau) = \sum_{\alpha}  {P}_{i\alpha}{P}_{j\alpha}\phi'(\langle V \rangle) \Theta\big(-\tau \big) e^{-|\tau|/\xi^{\alpha}_{\dot{\mathbf{n}}} }  
\end{align}
\begin{align}
    \label{eqn:APP_RVtn_eigen}
    {R}^{\tilde{\mathbf{V}}\dot{\mathbf{n}}}_{ij}(\tau) =  \sum_{\alpha}  {P}_{i\alpha}{P}_{j\alpha}\phi'(\langle V \rangle) \Theta\big(\tau \big) e^{-|\tau|/\xi^{\alpha}_{\dot{\mathbf{n}}} }  
\end{align}
\begin{align}
    \label{eqn:APP_RVnt_eigen}
    {R}^{\mathbf{V}\tilde{\mathbf{n}}}_{ij}(\tau) = \sum_{\alpha}  {P}_{i\alpha}{P}_{j\alpha} \lambda_\alpha  \Theta\big(-\tau \big)e^{-|\tau|/\xi^{\alpha}_{\dot{\mathbf{n}}}} 
\end{align}
\begin{align}
    \label{eqn:APP_RntV_eigen}
    {R}^{\tilde{\mathbf{n}}\mathbf{V}}_{ij}(\tau) = \sum_{\alpha}  {P}_{i\alpha}{P}_{j\alpha} \lambda_\alpha  \Theta\big(\tau \big)e^{-|\tau|/\xi^{\alpha}_{\dot{\mathbf{n}}}} 
\end{align}
\begin{align}
    \label{eqn:APP_RVxt_eigen}
    \begin{split}
        {R}^{\mathbf{V}\tilde{\mathbf{x}}}_{ij}(\tau) =& \sum_{\alpha}  {P}_{i\alpha}{P}_{j\alpha} \lambda_\alpha \left( \frac{1}{ \xi^{\alpha}_{\dot{\mathbf{n}}} } + \frac{1}{\xi^{\alpha}_{\mathbf{x}}} \right)^{-1} \left( \Theta\big(\tau \big)e^{-|\tau|/\xi^{\alpha}_{\dot{\mathbf{n}}}} + \Theta\big(-\tau \big)e^{-|\tau|/\xi^{\alpha}_{\mathbf{x}}} \right) 
    \end{split}
\end{align}
\begin{align}
    \label{eqn:APP_RxtV_eigen}
    \begin{split}
        {R}^{\tilde{\mathbf{x}}\mathbf{V}}_{ij}(\tau) =& \sum_{\alpha}  {P}_{i\alpha}{P}_{j\alpha} \lambda_\alpha \left( \frac{1}{ \xi^{\alpha}_{\dot{\mathbf{n}}} } + \frac{1}{\xi^{\alpha}_{\mathbf{x}}} \right)^{-1} \left( \Theta\big(-\tau \big)e^{-|\tau|/\xi^{\alpha}_{\dot{\mathbf{n}}}} + \Theta\big( \tau \big)e^{-|\tau|/\xi^{\alpha}_{\mathbf{x}}} \right)
    \end{split}
\end{align}
\begin{align}
    \label{eqn:APP_Rnnt_eigen}
    {R}^{\dot{\mathbf{n}}\tilde{\mathbf{n}}}_{ij}(\tau) = \sum_{\alpha} {P}_{i\alpha}{P}_{j\alpha} \Bigg[ \delta(\tau)+ \lambda_\alpha\phi'(\langle V \rangle) \Theta\big(-\tau \big) e^{-|\tau|/\xi^{\alpha}_{\dot{\mathbf{n}}}}  \Bigg]
\end{align}
\begin{align}
    \label{eqn:APP_Rntn_eigen}
    \begin{split}
        {R}^{\tilde{\mathbf{n}}\dot{\mathbf{n}}}_{ij}(\tau) =& \sum_{\alpha} {P}_{i\alpha}{P}_{j\alpha} \Bigg[ \delta(\tau)+ \lambda_\alpha\phi'(\langle V \rangle) \Theta\big(\tau \big) e^{-|\tau|/\xi^{\alpha}_{\dot{\mathbf{n}}}}  \Bigg]
    \end{split}
\end{align}
\begin{align}
    \label{eqn:APP_Rnxt_eigen}
    \begin{split}
        {R}^{\dot{\mathbf{n}}\tilde{\mathbf{x}}}_{ij}(\tau) =& \sum_{\alpha} {P}_{i\alpha}{P}_{j\alpha}\phi'(\langle V \rangle) \left( \frac{1}{ \xi^{\alpha}_{\dot{\mathbf{n}}} } + \frac{1}{\xi^{\alpha}_{\mathbf{x}}} \right)^{-1} \Bigg[ \Theta\big(\tau  \big)  e^{-|\tau|/\xi^{\alpha}_{\mathbf{x}}} +\Theta\big(-\tau \big)e^{-|\tau|/\xi^{\alpha}_{\dot{\mathbf{n}}}}  \Bigg]
    \end{split}
\end{align}
\begin{align}
    \label{eqn:APP_Rxtn_eigen}
    \begin{split}
        {R}^{\tilde{\mathbf{x}}\dot{\mathbf{n}}}_{ij}(\tau) =& \sum_{\alpha} {P}_{i\alpha}{P}_{j\alpha}\phi'(\langle V \rangle) \left( \frac{1}{ \xi^{\alpha}_{\dot{\mathbf{n}}} } + \frac{1}{\xi^{\alpha}_{\mathbf{x}}} \right)^{-1} \Bigg[ \Theta\big(-\tau  \big)  e^{-|\tau|/\xi^{\alpha}_{\mathbf{x}}} +\Theta\big(\tau \big)e^{-|\tau|/\xi^{\alpha}_{\dot{\mathbf{n}}}}  \Bigg]
    \end{split}
\end{align}
\begin{align}
    \label{eqn:APP_Rxxt_eigen}
    {R}^{\mathbf{x}\tilde{\mathbf{x}}}_{ij}(\tau) = \sum_{\alpha} {P}_{i\alpha}{P}_{j\alpha} \Theta\big(\tau \big) e^{-|\tau|/\xi^{\alpha}_{\mathbf{x}}}
\end{align}
\begin{align}
    \label{eqn:APP_Rxtx_eigen}
    {R}^{\tilde{\mathbf{x}}\mathbf{x}}_{ij}(\tau) = \sum_{\alpha} {P}_{i\alpha}{P}_{j\alpha} \Theta\big(-\tau \big) e^{-|\tau|/\xi^{\alpha}_{\mathbf{x}}}
\end{align}


\end{widetext}

\newpage

\end{document}